\documentclass[
	aps, aps,prd,longbibliography, superscriptaddress, twocolumn,
	10pt
	floatfix, 
    nofootinbib,
	tightenlines
]{revtex4-1}
\usepackage[final]{graphicx}
\usepackage{times,bbm,amsmath,amssymb}
\usepackage{epsfig,color}
\usepackage{xcolor}
\usepackage{hyperref}
\hypersetup{
    colorlinks = true
}
\usepackage{cleveref}
\usepackage{microtype}

\usepackage{float,siunitx}
\usepackage[caption = false]{subfig}

\usepackage[greek,english]{babel}
\usepackage{thumbpdf,enumerate}
\usepackage{booktabs}
\usepackage{sidecap}
\usepackage[scaled=.8]{couriers}
\usepackage{multirow}
\usepackage{placeins}
\usepackage{relsize}
\usepackage{pst-grad,bm}
\usepackage{epigraph}
\usepackage{gensymb}
\usepackage{longtable}
\usepackage{ulem} 
\usepackage{acronym}
\usepackage{physics}
\usepackage{easyReview}

\usepackage[columnwise]{lineno}

\DeclareUnicodeCharacter{0301}{\'{e}}

\begin{document}


\vspace{10pt}


\title{Interferometric imaging of amplitude and phase of spatial biphoton states}

\author{Danilo Zia}
\address{Dipartimento di Fisica, Sapienza Universit\`{a} di Roma, Piazzale Aldo Moro 5, I-00185 Roma, Italy}

\author{Nazanin Dehghan $^\dagger$}
\address{Nexus for Quantum Technologies, University of Ottawa, Ottawa, K1N 6N5, ON, Canada}

\author{Alessio D'Errico $^\dagger$} 
\email{aderrico@uottawa.ca}
\address{Nexus for Quantum Technologies, University of Ottawa, Ottawa, K1N 6N5, ON, Canada}
\author{Fabio Sciarrino}
\address{Dipartimento di Fisica, Sapienza Universit\`{a} di Roma, Piazzale Aldo Moro 5, I-00185 Roma, Italy}

\author{Ebrahim Karimi}
\address{Nexus for Quantum Technologies, University of Ottawa, Ottawa, K1N 6N5, ON, Canada}
\affiliation{National Research Council of Canada, 100 Sussex Drive, K1A 0R6, Ottawa, ON, Canada}

\begin{abstract}
High-dimensional biphoton states are promising resources for quantum applications, ranging from high-dimensional quantum communications to quantum imaging. A pivotal task is fully characterising these states, which is generally time-consuming and not scalable when projective measurement approaches are adopted. However, new advances in coincidence imaging technologies allow for overcoming these limitations by parallelising multiple measurements. Here, we introduce biphoton digital holography, in analogy to off-axis digital holography, where coincidence imaging of the superposition of an unknown state with a reference one is used to perform quantum state tomography. We apply this approach to single photons emitted by spontaneous parametric down-conversion in a nonlinear crystal when the pump photons possess various quantum states. The proposed reconstruction technique allows for a more efficient (3 order-of-magnitude faster) and reliable (an average fidelity of 87\%)  characterisation of states in arbitrary spatial modes bases, compared with previously performed experiments. Multi-photon digital holography may pave the route toward efficient and accurate computational ghost imaging and high-dimensional quantum information processing.
\end{abstract}

\maketitle 
\def\thefootnote{$\dagger$}\footnotetext{These authors contributed equally to this work}\def\thefootnote{\arabic{footnote}}

\section{Introduction}

Photonic qudits are emerging as an essential resource for environment-resilient quantum key distribution~\cite{cerf2002security,sheridan2010security,ding2017high,Sit17,bouchard2018experimental}, quantum simulation~\cite{aspuru2012photonic, d2021quantum} and quantum imaging and metrology~\cite{erhard2020advances, undetected2022, polino2020photonic}. The availability of unbounded photonic degrees of freedom, such as time-bins~\cite{flamini2018photonic}, temporal modes~\cite{ansari2018tailoring}, orbital angular momentum and radial number~\cite{allen_0AM_1992,erhard2018twisted,d2021quantum}, allows for encoding large amounts of information in fewer photons than would be required by qubit-based protocols (e.g., when using only polarization). At the same time, the large dimensionality of these states, such as those emerging from the generation of photon pairs, poses an intriguing challenge for what concerns their measurement. The number of projective measurements necessary for a full-state tomography scales exponentially with the dimensionality of the Hilbert space under consideration~\cite{thew2002qudit,nielsen2002quantum,eisert2020quantum,agnew2011tomography}. This issue can be tackled  with adaptive tomographic approaches \cite{huszar2012adaptive,mahler2013adaptive, rambach2021robust} or compressive techniques \cite{gross2010quantum, bouchard2019compressed}, which are, however, constrained by \textit{a priori} hypotheses on the quantum state under study. Note that quantum state tomography via projective measurement becomes challenging when the dimension of the quantum state is not a power of a prime number~\cite{bent2015experimental}. Here, we try to tackle the tomographic challenge, in the specific contest of spatially correlated biphoton states, looking for an interferometric approach inspired by digital holography~\cite{leith1964wavefront,yamaguchi2006phase,verrier2011off, d2017measuring, fu2020universal,ariyawansa2021amplitude}, familiar in classical optics. We show that the coincidence imaging of the superposition of two biphoton states, one unknown and one used as a reference state, allows retrieving the spatial distribution of phase and amplitude of the unknown biphoton wavefunction. Coincidence imaging can be achieved with modern EMCCD cameras~\cite{brida2010experimental,bolduc2017acquisition}, SPAD arrays~\cite{unternahrer2018super,zarghami202032,eckmann2020characterization} or time stamping cameras~\cite{fisher2016timepixcam, nomerotski2019imaging, nomerotski2023intensified}. These technologies are commonly exploited in quantum imaging, such as ghost imaging experiments~\cite{moreau2019imaging, zhang2020multidimensional,defienne2019quantum,salari2021quantum} or quantum superresolution \cite{tenne2019super,toninelli2019resolution,defienne2022pixel}, as well as for fundamental applications, such as characterizing two-photon correlations \cite{bolduc2017acquisition,boucher2021engineering}, imaging of high-dimensional Hong-Ou-Mandel interference~\cite{devaux2020imaging, zhang2021high,gao2022high,zhang2022ray}, and visualization of the violation of Bell inequalities~\cite{moreau2019imagingbell}. Holography techniques have been recently proposed in the context of quantum imaging~\cite{defienne2021polarization,topfer2022quantum,thekkadath2023intensity}; demonstrating the phase-shifting digital holography in a coincidence imaging regime using polarization entanglement~\cite{defienne2021polarization}, and exploiting induced coherence, i.e. the reconstruction of phase objects through digital holography of undetected photons~\cite{topfer2022quantum}.
\begin{figure*}[t]
    \centering
    \includegraphics[width=\textwidth]{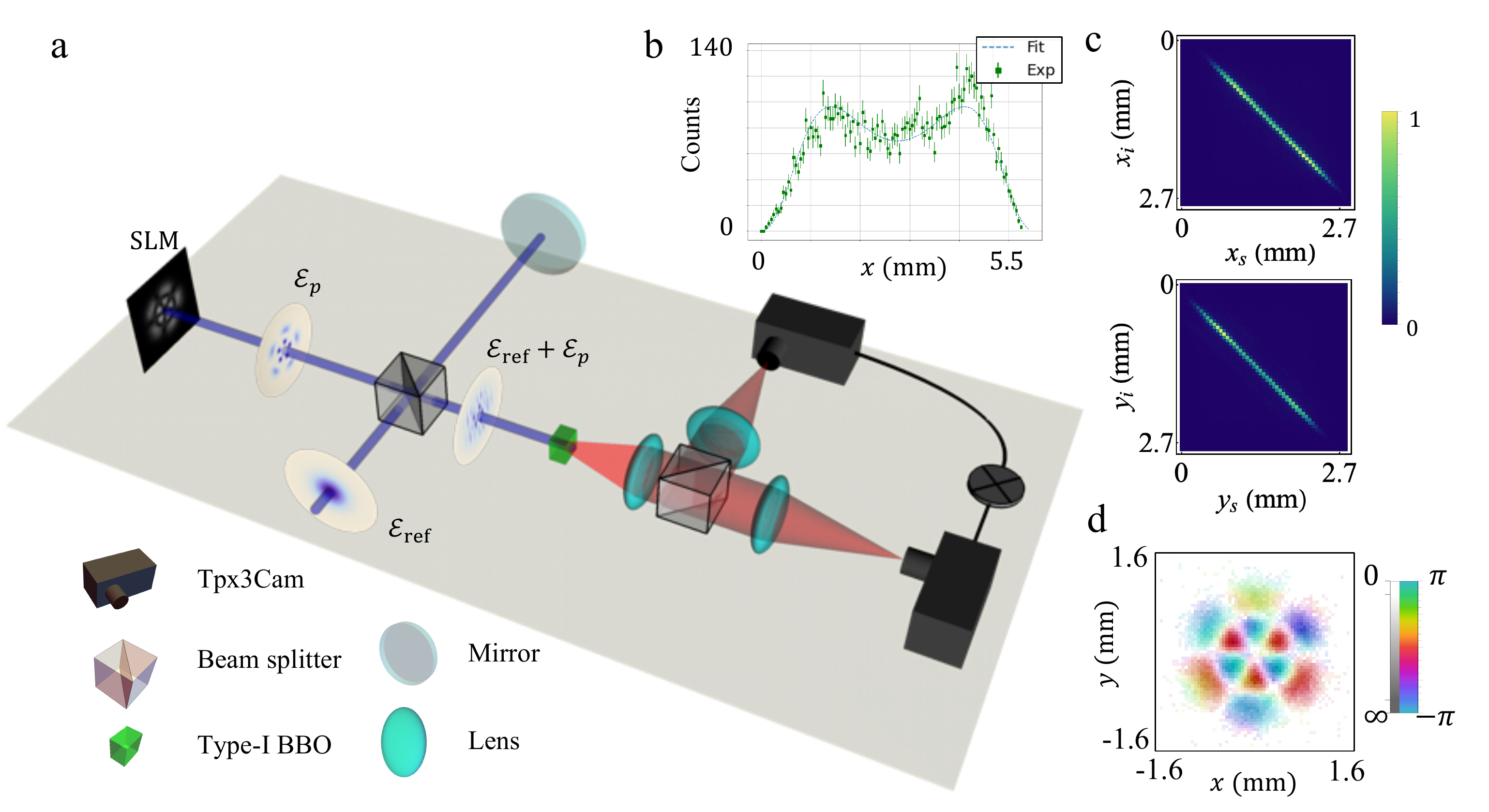}
    \caption{\textbf{Biphoton state holographic reconstruction:} a) Sketch of the experimental setup: a 405 nm laser in a Gaussian mode ($\mathcal{E}_{\text{ref}}$) enters a Michelson interferometer, where an Ultraviolet Spatial Light Modulator (UV-SLM) in one arm is used to shape and generate the unknown pump field ($\mathcal{E}_{p}$). The interferometer's output is the superposition of the reference and unknown pump field, which is then shined on a 0.5-mm-thick Type-I BBO crystal. Photon pairs are consequently generated and, after being separated into identical copies, sent on single photon sensor arrays. b) By placing the camera in the far field of the crystal and pumping with a large Gaussian beam, we can reconstruct the phase matching function $A\text{sinc}(\alpha |\mathbf{q}_i-\mathbf{q}_s|^2+\zeta)$ by direct imaging. The figure shows a scan of the phase matching function with a nonlinear field yielding $A = 93 \pm 2$ counts,  $\alpha= (9.1 \pm 0.2) \times 10^{-6} $ mm$^{2}$ and $\zeta= 0.30 \pm 0.02$. c) Experimental correlations in the $x$ and $y$ coordinates obtained by placing the sensors in the image plane of the crystal. d) Example of reconstructed phase and amplitude of a biphoton state (represented in inverted HSV colours) when pumping the crystal with a superposition of LG modes: $\text{LG}_{1,3}+\text{LG}_{1,-3}$.}
    \label{fig:Exp_Setup}
\end{figure*}

In this work, we focus on the specific problem of reconstructing the quantum state, in the transverse coordinate basis, of two photons emerging from degenerate Spontaneous Parametric Down Conversion (SPDC). These states are characterised by strong correlations in the transverse position (considered on the plane where the two-photon generation happens), which can be observed in other kinds of photon sources, such as cold atoms \cite{parniak2017wavevector}. In these sources, the two-photon wavefunction strongly depends on the shape of the pump laser used to induce the down-conversion process \cite{walborn2010spatial}. The most used approach to reconstruct the biphoton state emitted by a nonlinear crystal in literature is based on projective techniques \cite{mair2001entanglement,agnew2011tomography, d2021full, zhang2014radial, salakhutdinov2012full}. This method has drawbacks for what concern measurement times since it needs successive measurements on non-orthogonal bases and for the signal loss due to diffraction. We proposed an imaging-based procedure capable of overcoming both issues mentioned above, while giving the full state reconstruction of the unknown state. The core idea lies in assuming the SPDC state induced by a plane wave as known, and in superimposing this state with the unknown biphoton one. Coincidence imaging allows, at the same time, bench-marking the existence of strong spatial correlations, removing background counts, and reconstructing, from a single image, that exhibits interference between the two states, the full two-photon wavefunction. We demonstrate this technique for pump beams in different spatial modes, such as Laguerre-Gaussian (LG) and Hermite-Gaussian (HG) modes. We investigate several physical effects from the reconstructed states, such as orbital angular momentum conservation, the generation of high-dimensional Bell states, parity conservation, and radial correlations. Remarkably, we show how, from a simple measurement, one can retrieve information about two-photon states in arbitrary spatial mode bases without the efficiency and alignment issues that affect previously implemented projective characterisation techniques. Indeed, depending on the source brightness and on the required number of detection events, the measurement time can be of the order of tens of seconds, while the previously implemented projective techniques required several hours and were limited to the exploration of a small subspace of spatial modes. As a latter example, we give a proof of principle demonstration of the use of this technique for quantum imaging applications.

\begin{figure*}[!t]
    \centering
    \includegraphics[width=\textwidth]{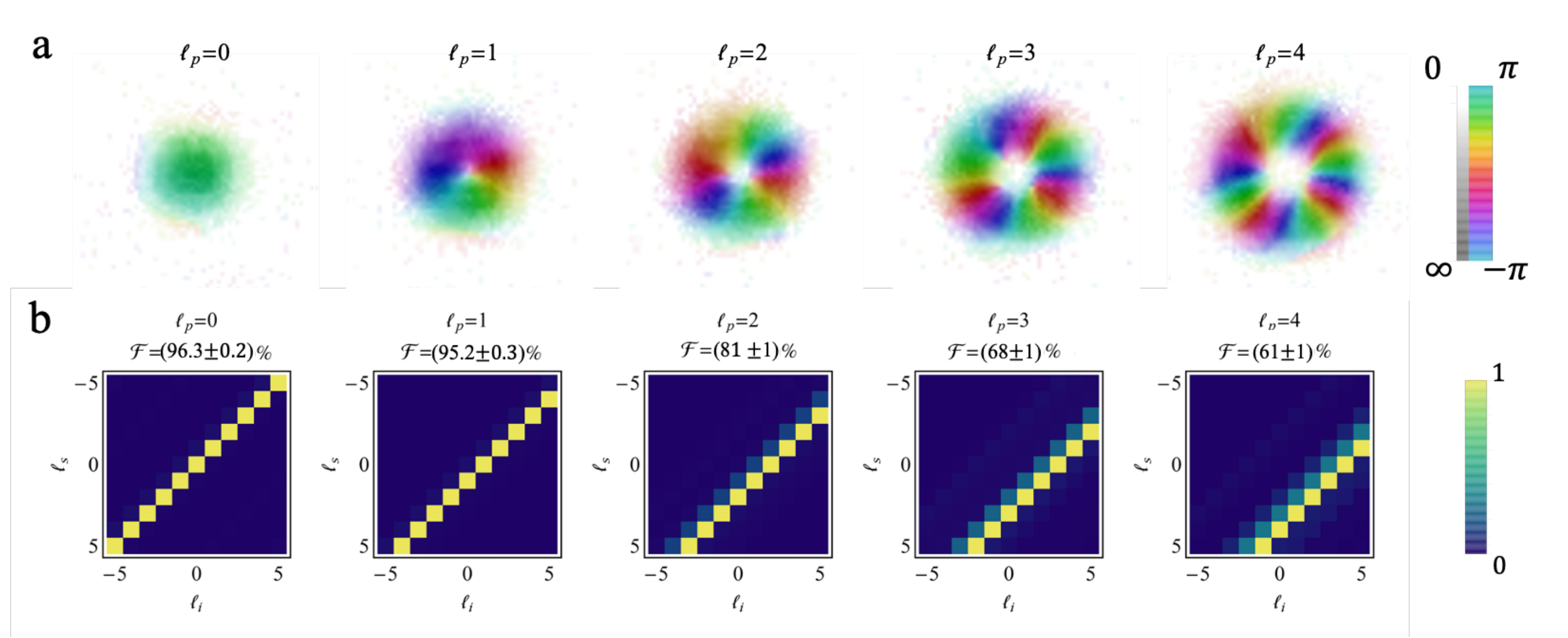}
    \caption{\textbf{SPDC photons OAM correlations.} Reconstructed field of the biphoton state for different OAM-carrying pump beams in $\text{LG}_{p_p=0,\ell_p}$ modes. (\textbf{a}) Shows the amplitude and phase of the state for different values of the pump OAM $\ell_p$. (\textbf{b}) Shows the OAM correlations density plots of generated SPDC photons. It can be seen how increasing the pump OAM, the sum of the OAM values for the idler and signal photons shift in agreement with the conservation law of Eq. \eqref{eq:OAM_cons}. The error analysis is outlined in the Methods.}
    \label{fig:OAM_conservation}
\end{figure*}

\section{Theory Background}
We consider the photon pair production in degenerate SPDC. In this process, a second-order nonlinear crystal, pumped by a laser beam with frequency $\omega_p$ and with a spatial amplitude $\mathcal{E}_p(x,y,z)$, produces (within a first-order approximation) photon pairs with frequencies $\omega_i$ and $\omega_s$ such that $\omega_p=\omega_i+\omega_s$, where the subscripts $s$ and $i$ denote the \textit{signal} and \textit{idler} photons, respectively. In the degenerate case, $\omega_i=\omega_s$, the two-photon state, written in the basis of transverse wavevector modes $\ket{\mathbf{q}}:=\ket{k_x,k_y}$, is given by~\cite{walborn2010spatial}:

\begin{align}\label{eq:spdc_momentum}
\ket{\Psi}=\mathcal{N}\iint&\,E_p(\mathbf{q}_i+\mathbf{q}_s)\,\text{sinc}(\alpha |\mathbf{q}_i-\mathbf{q}_s|^2+\zeta)\cr
\times&\ket{\mathbf{q}_i}\otimes\ket{\mathbf{q_s}}\,d^2{\mathbf{q}}_i\,d^2{\mathbf{q}}_s,
\end{align}
where $\mathcal{N}$ is a normalization factor, $E_p$ is the 2D Fourier transform of the pump mode on the crystal plane, $E_p(\mathbf{q})=\int \mathcal{E}_p(\bm{\rho},z=0)e^{i\bm{\rho}\cdot\mathbf{q} }d^2\bm{\rho}$, $\alpha=Lc/4\omega_p$ ($L$ is the crystal length and $c$ is the speed of light in the medium ) and $\zeta$ is the longitudinal mismatch which depends on the crystal orientation. The contribution of the sinc function is related to the phase matching of the SPDC process. It is interesting to consider the same state in the transverse coordinate basis $\ket{\bm{\rho}}=\int e^{-i\bm{\rho}\cdot\mathbf{q}}\ket{\bm{q}}d^2\bm{q}$:

\begin{align} \label{eq:spdc_position}
    \ket{\Psi}=\mathcal{N'}\iint &\mathcal{E}_p(\bm{\rho}_i+\bm{\rho}_s,z=0)\,{\cal FT}[\text{sinc}](\bm{\rho}_i-\bm{\rho}_s)\nonumber\\ \times&\ket{\bm{\rho}_i}\otimes\ket{\bm{\rho}_s}\,d^2\bm{\rho}_i\,d^2\bm{\rho}_s,
\end{align}
where constants have been included in the normalization factor $\mathcal{N}'$. Since the width of the Fourier transform of the phase matching function, ${\cal FT}[\text{sinc}]$, is of the order of $\alpha$, i.e. of the order of $\sqrt{L\lambda_p}$, for thin crystals, the SPDC state is well approximated by:

\begin{align} \label{eq:spdc_thincryst}
    \ket{\Psi}=\mathcal{N'}\int\mathcal{E}_p(2\bm{\rho}) \ket{\bm{\rho}}\otimes\ket{\bm{\rho}}\,d^2\bm{\rho}.
\end{align}
This result highlights the strong correlations in the transverse position basis and that the pump shape essentially determines the two-photon state. This suggests a simple way to reconstruct the two-photon state experimentally through an interferometric technique. Together with the unknown SPDC state, one can generate a reference two-photon state $\ket{\Psi_\text{ref}}$ with strong position correlations whose amplitude and phase of the biphoton wavefunction are known. The resulting state is $\ket{\Psi_\text{TOT}}:=\ket{\Psi}+\ket{\Psi_\text{ref}}$. When measuring position-dependent coincidences in the image plane of the crystal, one can obtain information about the phase of the unknown biphoton state. Indeed, in the thin crystal approximation, the diagonal contributions of the coincidence count rate are given by,
\begin{equation}\label{eq:coinc}
    \mathcal{C}(\bm{\rho},\bm{\rho}):=|\braket{\bm{\rho},\bm{\rho}}{\Psi_\text{TOT}}|^2=|\mathcal{E}_p(2\bm{\rho})+\mathcal{E}_\text{ref}(2\bm{\rho})|^2.
\end{equation}
Here, $\mathcal{E}_\text{ref}$ is the pump shape used to generate the reference SPDC state. Ideally, $\mathcal{E}_\text{ref}$ can be a plane wave or, in practice, a Gaussian beam with a large waist. By controlling the reference pump beam one can map to the two-photon case any interferometric technique that is used in classical optics for amplitude and phase reconstruction. In this work, we experimentally implemented off-axis digital holography, where the reference beam is a Gaussian beam with a tilted wavefront. The proposed scheme can be implemented in two measurement steps: first, the correlations in the crystal image plane are measured to confirm the validity of the thin crystal approximation, second, the coincidences corresponding to Eq.~\eqref{eq:coinc} are evaluated and the biphoton state extracted from the resulting interference pattern. Beyond the thin crystal approximation limit, the effect of the phase-matching function is to reduce the visibility of the interference fringes. In such a scenario, the pump field can still be reconstructed, but it will not correspond to the two-photon wavefunction. The latter can be reconstructed by additionally measuring the phase matching function, whose amplitude can be easily reconstructed in the far field since imaging the state in this plane corresponds to performing a Fourier transform on it.

\section{Experimental setup and results}
\begin{figure*}[ht!]
\centering\includegraphics[width=\textwidth]{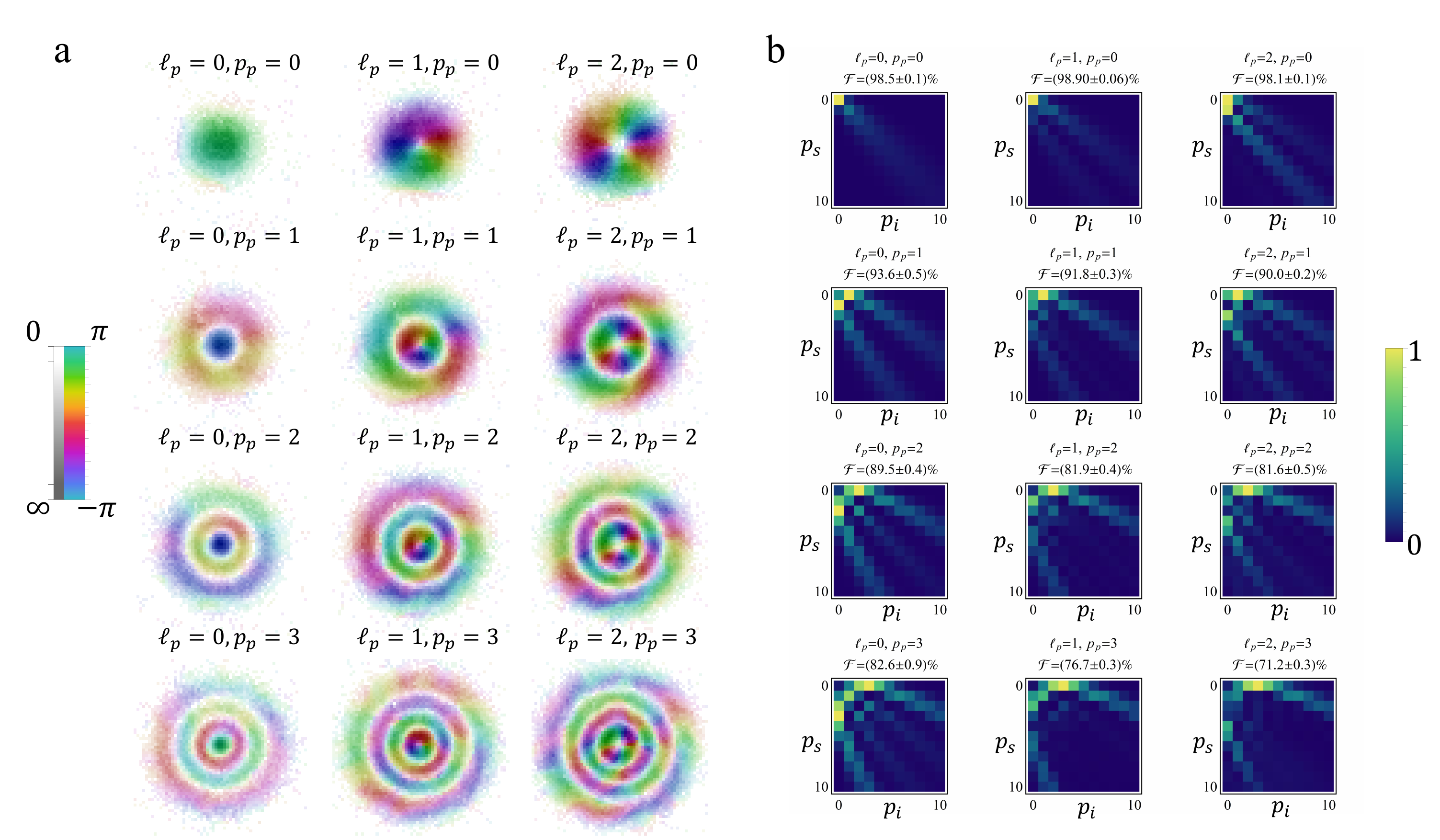}
    \caption{\textbf{Radial mode correlations.} (\textbf{a}) reconstructed biphoton fields obtained by pumping the crystal with LG modes. (b) Shows the correlations in the radial number $p$. In all the plots the OAM of signal and idler has been fixed to $\ell_i=0, \ell_s=\ell_p$. The Fidelities are obtained assuming the theoretical state calculated in the thin crystal approximation. The error analysis is reported in the Methods.}
    \label{fig:p_correlations}
\end{figure*}
Following the theoretical description of the previous section, we experimentally implemented a platform in which, through off-axis digital holography, the biphoton state, emitted via SPDC by a Type I crystal of $\beta-$Barium Borate (BBO), is reconstructed. In this proof of principle experiment, we generate the unknown and reference SPDC states in the same crystal. A visual scheme of the setup is reported in Fig. \ref{fig:Exp_Setup}-a  (see Methods for details).
For our purpose, we built a Michelson interferometer placing a Spatial Light Modulator (SLM) in one of the two arms. This allows creating a pump beam in the mode $\mathcal{E}_p+\mathcal{E}_\text{ref}$, where the reference mode is a wide Gaussian with a tilted wavefront $\mathcal{E}_\text{ref}=\exp(-r^2/w_r^2)\exp(i 2\pi(x+y)/\Lambda)$. The mean transverse momentum $2\pi/\Lambda$ is chosen to maximize the spatial resolution of the reconstructed field, and $w_r$ is chosen to be larger than the characteristic waist parameter of $\mathcal{E}_p$, denoted as $w_p$. The interferometer's output is sent on the BBO crystal, and the two photons' state is recorded using a time-stamping camera (TPX3CAM). The camera is composed of a matrix of $256\times256$ time-stamping pixels of 55 $\mu$m size and with $\approx 1$ ns time-resolution. We collected data for both the crystal's Fourier and the image plane. The Fourier plane data (acquired for an input Gaussian beam) is exploited to characterise the phase-matching function (see Figure \ref{fig:Exp_Setup}-b) that gives the main contribution to the state wavefunction in this plane. The second one, instead, is used to reconstruct the bi-photon state in the thin-crystal approximation. In each case, we split the emitted photons by imaging two copies of the SPDC into different regions of the camera sensor, allowing one to check for coincidences between different pixels. By observing the spatial correlations, we could verify the correctness of the thin-crystal approximation. This is shown by the sharp, $\approx$1-pixel wide, spatial correlations observed in all the cases under analysis; an example is reported in Fig. \ref{fig:Exp_Setup}-c. Data were collected in 1 minute of exposure for each spatial mode under analysis. In particular, we collected both the interference pattern between the two states and a coincidence image of down-converted light induced by $\mathcal{E}_{p}$ only. The first one is used to retrieve the phase of the state under analysis, while the second one already gives the amplitude of the biphoton field. By exploiting this reconstruction, we were able to fully characterise the biphoton state. An example of the reconstructed phase and amplitude of the biphoton state is in Fig \ref{fig:Exp_Setup}-d. Moreover, we also characterise the amplitude of the phase-matching function. To do so, we collect an image of the far field by placing an additional lens in front of the TPX3CAM camera and entering the crystal with a wide Gaussian beam. Fitting the collected data with the sinc function (Eq.~\eqref{eq:spdc_momentum}), we obtained $\alpha= (9.1 \pm 0.2) \times 10^{-6}$ mm$^{2}$ and $\zeta= 0.30 \pm 0.02$. From $\alpha$, we retrieve a value of the crystal length $L_{exp} = 0.56 \pm 0.01$ mm, which is in very good agreement with the nominal value $L_{nom} = 0.5$ mm. The phase-matching fit is shown in figure Fig.~\ref{fig:Exp_Setup}-b. 

Once the bi-photon state is given, one can extract any desired information about this state, for example, correlations in different degrees of freedom, entanglement, and the decomposition in arbitrary sets of spatial modes. 

\begin{figure*}[ht!]
\centering\includegraphics[width=\textwidth]{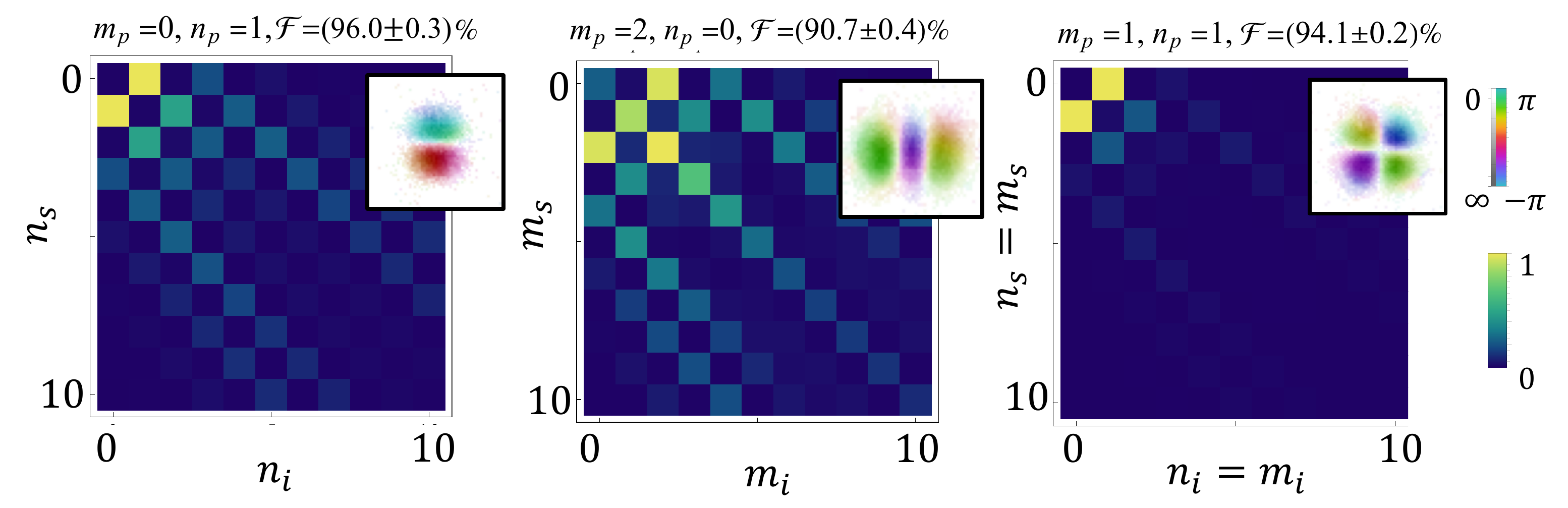}
    \caption{\textbf{Hermite-Gauss correlations.} When pumping the crystal with $\mathcal{E}_p(x,y)=\text{HG}_{m,n}(x,y)$ we observe biphoton correlations in the basis of HG modes which highlight the parity conservation of the SPDC process. Insets show the reconstructed biphoton fields from which the correlations have been extracted. Upon each plot the Fidelity between the retrieved field and the theoretical one in the thin crystal approximation is reported. The error analysis is discussed in the Methods.}
    \label{fig:hgcorr}
\end{figure*}

One of the degrees of freedom of light, which has been extensively studied for high-dimensional quantum applications, is the Orbital Angular Momentum (OAM). Modes in OAM eigenstates are represented by wavefunctions possessing a phase term of the form $\braket{\phi}{\ell}:=\exp(i\ell \phi)$ in position representation. Here, $\phi$ is the azimuthal angle in cylindrical coordinates, and $\ell\in\mathbb{Z}$ is the OAM value (along the propagation direction) in units of $\hbar$ carried by a photon in such a state. In the following, a Gaussian radial wavefunction will be assumed $\braket{r,\phi}{\ell}:=\exp(-r^2/w_p^2)\exp(i\ell \phi)$, with waist equal to the waist parameter in $\mathcal{E}_p$.

This choice (which corresponds to the waist plane expression of Hypergeometric-Gaussian modes \cite{karimi_07}) allows for finite radial integrals when considering the SPDC decomposition in OAM modes:  $\ket{\Psi}=\sum_{\ell_i,\ell_s}c_{\ell_i,\ell_s}\ket{\ell_i}\otimes\ket{\ell_s}$, where $\ell_{i}, \ell_{s}$ are respectively the azimuthal indices of the idler and signal photons, and the coefficients of the decomposition are,
\begin{equation}\label{eq:oamcoeff}
    c_{\ell_i,\ell_s}=\iint \mathcal{E}_p(r,\phi)\text{e}^{-2r^2/w_p^2}\,\text{e}^{-i(\ell_i+\ell_s)\phi}\,rdr\,d\phi.
\end{equation}
In particular, when the pump beam carries an OAM equal to $\ell_{p}$ one has the OAM conservation law:
\begin{equation}    \label{eq:OAM_cons}
    \ell_{p}=\ell_{i}+\ell_{s},
\end{equation}
which can be immediately deduced from \eqref{eq:oamcoeff}~\cite{walborn2010spatial} and was first demonstrated in~ ref.~\cite{mair2001entanglement}. 
We investigated this relationship for several OAM values by entering the crystal with LG modes \cite{siegman1986lasers}. These are a set of modes defined as: $LG_{p,\ell}(r,\phi):=\braket{r,\phi}{p,\ell}\propto(r/w)^{\abs{\ell}}L_{p}^{\abs{\ell}}(2r^2/w^2)\exp(-(r/w)^2)\exp(i\ell\phi)$, where $L_{p}^{\abs{\ell}}(x)$ are associated Laguerre polynomials. They are cylindrically symmetric modes carrying OAM and with minimal divergence in free space~\cite{vallone2016general}. We analyzed the case in which we entered the crystal with states having azimuthal index $\ell \in \{0,1,2,3,4\}$, the results are reported in Fig.~\ref{fig:OAM_conservation}. It is evident how increasing the OAM carried by the pump causes the OAM correlations to shift in agreement with Eq.~\eqref{eq:OAM_cons} (as also observed in, e.g. \cite{mair2001entanglement}). It has been observed that the SPDC state in the OAM basis can be approximated as a high-dimensional Bell state~\cite{agnew2011tomography,dada2011experimental}. In Fig.~\ref{fig:OAM_conservation}-b, we also report the Fidelities between the reconstructed states and a high-dimensional Bell state $\ket{\Psi_\text{Bell}}:=1/\sqrt{(2L+1-\ell_p)}\sum_{\ell=-L+\ell_p,L}\ket{\ell}\ket{\ell_p-\ell}$, with $L=5$. As noted in ref.~\cite{agnew2011tomography} these values of Fidelities indicate how the reconstructed state can violate high-dimensional Bell inequalities. Fidelity values decrease with a higher $\ell_p$ mainly due to imperfections in the pump preparations (hence the pump beam is better described as a superposition of OAM states).

The radial index $p\in\mathbb{N}$ of LG modes corresponds to the number of radial zeros and can be treated as a quantum number~\cite{PhysRevA.89.063813,zhang2018violation}. In Fig. \ref{fig:p_correlations}-a, experimental results of biphoton states for pump beams, prepared as LG modes, are shown. From the reconstructed states, the coefficients of the SPDC state decomposition in LG modes $\ket{\Psi}=\sum_{p_i,\ell_i}^{p_s,\ell_s}C_{p_i,\ell_i}^{p_s,\ell_s}\ket{p_i,\ell_i}\otimes\ket{p_s,\ell_s}$ were extracted. Figure \ref{fig:p_correlations}-b shows experimental correlations in radial indexes (with OAM indexes of idler and signal fixed as $\ell_i=0, \ell_s=\ell_p$). When choosing the waist parameter of the decomposition to be equal to the pump waist, the correlations are maximized for $p_{i,s}=p_p,\, p_{s,i}=0$. This can be understood from the similarity between the integral expression of $C_{p_i,\ell_i}^{p_s,\ell_s}$ and the orthogonality relationship of LG modes. Within the considered subspaces (where indices $p_{i,s}$ are bounded from 0 to 10 and $\ell_i=0, \ell_s=\ell_p$), the Fidelity $\mathcal{F}=|\sum_{p_i}^{p_s}{C^*}_{p_i,0}^{p_s,\ell_p}C^{th}{}^{p_s,\ell_p}_{p_i,0}|^2$ was evaluated, where ${C^*}_{p_i,0}^{p_s,\ell_p}$ are the measured coefficients and $C^{th}{}^{p_s,\ell_p}_{p_i,0}$ are the ones expected from the thin crystal approximation. 
In this approximation, the spatial modes' parity is also conserved \cite{walborn2005conservation}. This effect can be highlighted by considering pump beams in HG modes~\cite{siegman1986lasers}, which, on the crystal plane, read $HG_{m_p,n_p}(x,y):=\braket{x,y}{m_p, n_p}\propto \exp(-(x^2+y^2)/w_p^2)h_{m_p}(x/w_p)h_{n_p}(y/w_p)$, where $h_m(x)$ are Hermite polynomials of order $m$. HG modes form a complete, orthonormal set with even or odd functions along the $x$ or $y$ directions. Thus, the SPDC state can be decomposed as $\ket{\Psi}=\sum_{m_i,n_i}^{m_s,n_s}C_{m_i,n_i}^{m_s,n_s}\ket{m_i,n_i}\otimes\ket{m_s,n_s}$. This basis has been extensively studied \cite{walborn2005conservation,zhang2016hong, miatto2012spatial} and recently considered for biphoton super-resolution measurements \cite{grenapin2022super}. When studying the SPDC correlations on the basis of HG modes (chosen with the same waist parameter $w=w_p$ of the pump), one has the conservation laws: $n_p=\text{mod}(n_i+n_s, 2)$ and $m_p=\text{mod}(m_i+m_s, 2)$, as can be directly inferred from the parity of the integrands appearing in the expression of $C_{m_i,n_i}^{m_s,n_s}$ (see Methods and Ref. \cite{walborn2005conservation} for detailed proof). Figure~\ref{fig:hgcorr} shows calculated HG mode correlations for different states (with reconstructed biphoton amplitude and phase shown in the insets). The results show an excellent agreement with the theory; in particular, the parity conservation is evident from the chessboard-like correlation patterns.

\begin{figure}[ht!]
    \centering
    \includegraphics[width=\columnwidth]{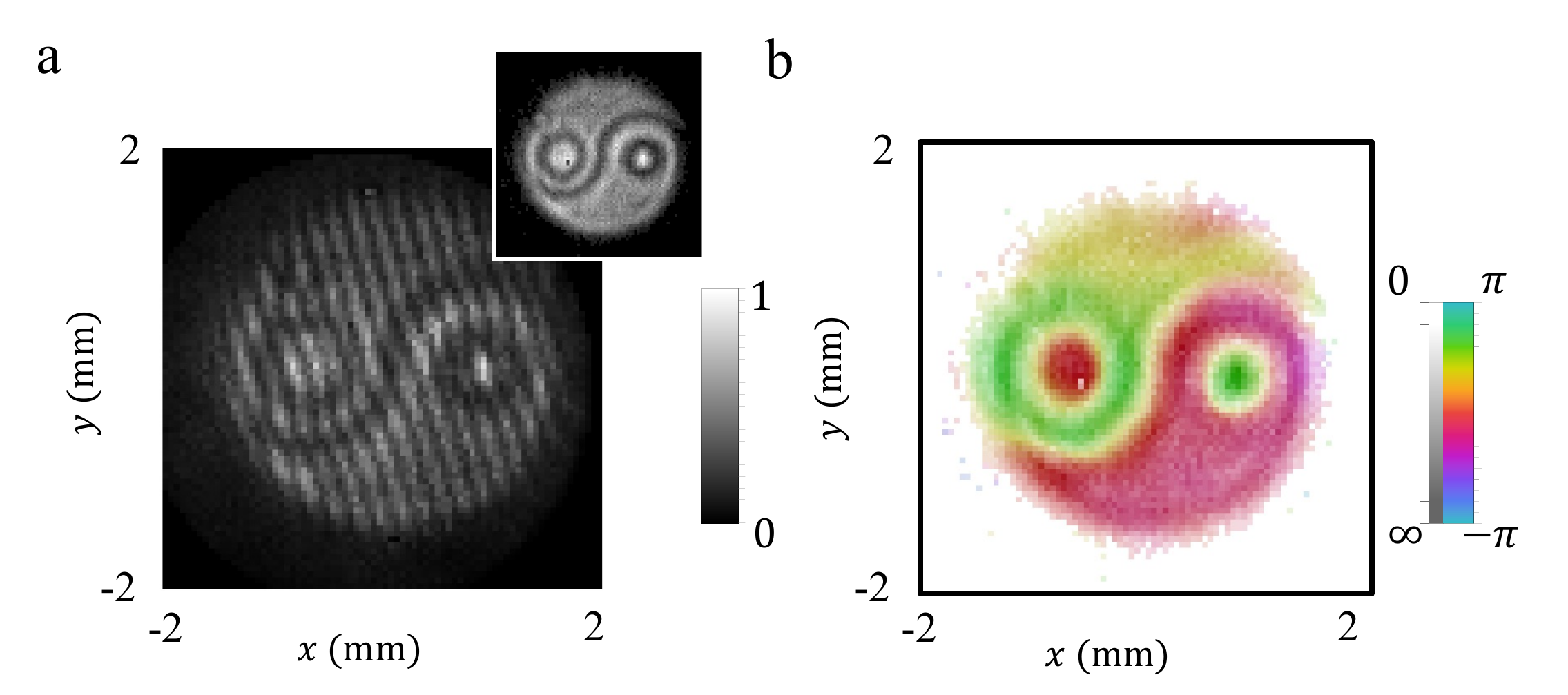}
    \caption{\textbf{Image reconstruction.} (\textbf{a}) Coincidence image of interference between a reference SPDC state and a state obtained by a pump beam with the shape of a Ying and Yang symbol (shown in the inset). The inset scale is the same as in the main plot. (\textbf{b}) Reconstructed amplitude and phase structure of the image imprinted on the ``unknown'' pump.}
    \label{fig:yinyang}
\end{figure}
Finally, figure \ref{fig:yinyang} shows an example of the potential applications of biphoton digital holography.
The unknown pump beam can carry information about an image or be scattered by a three-dimensional object. The information about the scatterer is transferred to the SPDC state and can be retrieved through our technique (Fig. \ref{fig:yinyang}-b). We show this in the case of off-axis holography, which can present limitations for complex structures due to the limited camera resolution. These limitations are not related to our proposal and can be improved by employing other approaches, e.g. on-axis phase-shifting digital holography \cite{yamaguchi2006phase}.

\section{Conclusion}
In this work, we introduced a novel approach for reconstructing the spatial structure of correlated two photons states. Our proposal exploits the strong spatial correlations typical of SPDC processes, the coherent superposition of two SPDC states and the possibility of imaging the amplitude of this superposition with a time-stamping camera. The experimental results showed how, from a single measurement, it is possible to retrieve, in postprocessing, a large amount of information about a two-photon spatial state, e.g. correlations in different degrees of freedom, entanglement and spatial mode decomposition in arbitrary bases. We gave examples analysing OAM and parity conservation, high-dimensional Bell states, and radial mode correlations. The results show the superiority of this technique, compared to projective techniques (e.g. the ones in refs. \cite{zhang2018violation,d2021full, valencia2021entangled}), in the context of benchmarking highly correlated quantum states. We achieved an enhancement on the reconstruction time up to 3 order of magnitude with high Fidelities for the biphoton states, obtaining an average Fidelity equal to $87 \%$. The lowest Fidelity values are due to imperfect pump preparations or an undesired spatially varying phase in the reference beam and not to intrinsic limitations of the technique. 
Although our proposal is based on a specific kind of two-photon state, it should be possible to generalise it to states where the correlations are not as sharp as in our case. A key ingredient is to generate reference states with spatial correlations that overlap well with the correlations in the unknown state. Hence, future investigations will be devoted to the generalisation of this approach to arbitrary two and multi-photon states. Moreover, we point out that the same technique presented here can be applied to measure biphoton states in the time-frequency degrees of freedom. Besides the quantum state reconstruction, future investigations will be devoted to the generalisation of our protocol to imaging experiments.

\bibliography{mrefs.bib}

\begin{thebibliography}{72}%
\makeatletter
\providecommand \@ifxundefined [1]{%
 \@ifx{#1\undefined}
}%
\providecommand \@ifnum [1]{%
 \ifnum #1\expandafter \@firstoftwo
 \else \expandafter \@secondoftwo
 \fi
}%
\providecommand \@ifx [1]{%
 \ifx #1\expandafter \@firstoftwo
 \else \expandafter \@secondoftwo
 \fi
}%
\providecommand \natexlab [1]{#1}%
\providecommand \enquote  [1]{``#1''}%
\providecommand \bibnamefont  [1]{#1}%
\providecommand \bibfnamefont [1]{#1}%
\providecommand \citenamefont [1]{#1}%
\providecommand \href@noop [0]{\@secondoftwo}%
\providecommand \href [0]{\begingroup \@sanitize@url \@href}%
\providecommand \@href[1]{\@@startlink{#1}\@@href}%
\providecommand \@@href[1]{\endgroup#1\@@endlink}%
\providecommand \@sanitize@url [0]{\catcode `\\12\catcode `\$12\catcode
  `\&12\catcode `\#12\catcode `\^12\catcode `\_12\catcode `\%12\relax}%
\providecommand \@@startlink[1]{}%
\providecommand \@@endlink[0]{}%
\providecommand \url  [0]{\begingroup\@sanitize@url \@url }%
\providecommand \@url [1]{\endgroup\@href {#1}{\urlprefix }}%
\providecommand \urlprefix  [0]{URL }%
\providecommand \Eprint [0]{\href }%
\providecommand \doibase [0]{http://dx.doi.org/}%
\providecommand \selectlanguage [0]{\@gobble}%
\providecommand \bibinfo  [0]{\@secondoftwo}%
\providecommand \bibfield  [0]{\@secondoftwo}%
\providecommand \translation [1]{[#1]}%
\providecommand \BibitemOpen [0]{}%
\providecommand \bibitemStop [0]{}%
\providecommand \bibitemNoStop [0]{.\EOS\space}%
\providecommand \EOS [0]{\spacefactor3000\relax}%
\providecommand \BibitemShut  [1]{\csname bibitem#1\endcsname}%
\let\auto@bib@innerbib\@empty
\bibitem [{\citenamefont {Cerf}\ \emph {et~al.}(2002)\citenamefont {Cerf},
  \citenamefont {Bourennane}, \citenamefont {Karlsson},\ and\ \citenamefont
  {Gisin}}]{cerf2002security}%
  \BibitemOpen
  \bibfield  {author} {\bibinfo {author} {\bibfnamefont {Nicolas~J}\
  \bibnamefont {Cerf}}, \bibinfo {author} {\bibfnamefont {Mohamed}\
  \bibnamefont {Bourennane}}, \bibinfo {author} {\bibfnamefont {Anders}\
  \bibnamefont {Karlsson}}, \ and\ \bibinfo {author} {\bibfnamefont {Nicolas}\
  \bibnamefont {Gisin}},\ }\bibfield  {title} {\enquote {\bibinfo {title}
  {Security of quantum key distribution using d-level systems},}\ }\href@noop
  {} {\bibfield  {journal} {\bibinfo  {journal} {Physical review letters}\
  }\textbf {\bibinfo {volume} {88}},\ \bibinfo {pages} {127902} (\bibinfo
  {year} {2002})}\BibitemShut {NoStop}%
\bibitem [{\citenamefont {Sheridan}\ and\ \citenamefont
  {Scarani}(2010)}]{sheridan2010security}%
  \BibitemOpen
  \bibfield  {author} {\bibinfo {author} {\bibfnamefont {Lana}\ \bibnamefont
  {Sheridan}}\ and\ \bibinfo {author} {\bibfnamefont {Valerio}\ \bibnamefont
  {Scarani}},\ }\bibfield  {title} {\enquote {\bibinfo {title} {Security proof
  for quantum key distribution using qudit systems},}\ }\href@noop {}
  {\bibfield  {journal} {\bibinfo  {journal} {Physical Review A}\ }\textbf
  {\bibinfo {volume} {82}},\ \bibinfo {pages} {030301} (\bibinfo {year}
  {2010})}\BibitemShut {NoStop}%
\bibitem [{\citenamefont {Ding}\ \emph {et~al.}(2017)\citenamefont {Ding},
  \citenamefont {Bacco}, \citenamefont {Dalgaard}, \citenamefont {Cai},
  \citenamefont {Zhou}, \citenamefont {Rottwitt},\ and\ \citenamefont
  {Oxenl{\o}we}}]{ding2017high}%
  \BibitemOpen
  \bibfield  {author} {\bibinfo {author} {\bibfnamefont {Yunhong}\ \bibnamefont
  {Ding}}, \bibinfo {author} {\bibfnamefont {Davide}\ \bibnamefont {Bacco}},
  \bibinfo {author} {\bibfnamefont {Kjeld}\ \bibnamefont {Dalgaard}}, \bibinfo
  {author} {\bibfnamefont {Xinlun}\ \bibnamefont {Cai}}, \bibinfo {author}
  {\bibfnamefont {Xiaoqi}\ \bibnamefont {Zhou}}, \bibinfo {author}
  {\bibfnamefont {Karsten}\ \bibnamefont {Rottwitt}}, \ and\ \bibinfo {author}
  {\bibfnamefont {Leif~Katsuo}\ \bibnamefont {Oxenl{\o}we}},\ }\bibfield
  {title} {\enquote {\bibinfo {title} {High-dimensional quantum key
  distribution based on multicore fiber using silicon photonic integrated
  circuits},}\ }\href@noop {} {\bibfield  {journal} {\bibinfo  {journal} {npj
  Quantum Information}\ }\textbf {\bibinfo {volume} {3}},\ \bibinfo {pages}
  {1--7} (\bibinfo {year} {2017})}\BibitemShut {NoStop}%
\bibitem [{\citenamefont {Sit}\ \emph {et~al.}(2017)\citenamefont {Sit},
  \citenamefont {Bouchard}, \citenamefont {Fickler}, \citenamefont
  {Gagnon-Bischoff}, \citenamefont {Larocque}, \citenamefont {Heshami},
  \citenamefont {Elser}, \citenamefont {Peuntinger}, \citenamefont
  {G\"{u}nthner}, \citenamefont {Heim}, \citenamefont {Marquardt},
  \citenamefont {Leuchs}, \citenamefont {Boyd},\ and\ \citenamefont
  {Karimi}}]{Sit17}%
  \BibitemOpen
  \bibfield  {author} {\bibinfo {author} {\bibfnamefont {Alicia}\ \bibnamefont
  {Sit}}, \bibinfo {author} {\bibfnamefont {Fr\'{e}d\'{e}ric}\ \bibnamefont
  {Bouchard}}, \bibinfo {author} {\bibfnamefont {Robert}\ \bibnamefont
  {Fickler}}, \bibinfo {author} {\bibfnamefont {J\'{e}r\'{e}mie}\ \bibnamefont
  {Gagnon-Bischoff}}, \bibinfo {author} {\bibfnamefont {Hugo}\ \bibnamefont
  {Larocque}}, \bibinfo {author} {\bibfnamefont {Khabat}\ \bibnamefont
  {Heshami}}, \bibinfo {author} {\bibfnamefont {Dominique}\ \bibnamefont
  {Elser}}, \bibinfo {author} {\bibfnamefont {Christian}\ \bibnamefont
  {Peuntinger}}, \bibinfo {author} {\bibfnamefont {Kevin}\ \bibnamefont
  {G\"{u}nthner}}, \bibinfo {author} {\bibfnamefont {Bettina}\ \bibnamefont
  {Heim}}, \bibinfo {author} {\bibfnamefont {Christoph}\ \bibnamefont
  {Marquardt}}, \bibinfo {author} {\bibfnamefont {Gerd}\ \bibnamefont
  {Leuchs}}, \bibinfo {author} {\bibfnamefont {Robert~W.}\ \bibnamefont
  {Boyd}}, \ and\ \bibinfo {author} {\bibfnamefont {Ebrahim}\ \bibnamefont
  {Karimi}},\ }\bibfield  {title} {\enquote {\bibinfo {title} {High-dimensional
  intracity quantum cryptography with structured photons},}\ }\href {\doibase
  10.1364/OPTICA.4.001006} {\bibfield  {journal} {\bibinfo  {journal} {Optica}\
  }\textbf {\bibinfo {volume} {4}},\ \bibinfo {pages} {1006--1010} (\bibinfo
  {year} {2017})}\BibitemShut {NoStop}%
\bibitem [{\citenamefont {Bouchard}\ \emph {et~al.}(2018)\citenamefont
  {Bouchard}, \citenamefont {Heshami}, \citenamefont {England}, \citenamefont
  {Fickler}, \citenamefont {Boyd}, \citenamefont {Englert}, \citenamefont
  {S{\'a}nchez-Soto},\ and\ \citenamefont {Karimi}}]{bouchard2018experimental}%
  \BibitemOpen
  \bibfield  {author} {\bibinfo {author} {\bibfnamefont {Fr{\'e}d{\'e}ric}\
  \bibnamefont {Bouchard}}, \bibinfo {author} {\bibfnamefont {Khabat}\
  \bibnamefont {Heshami}}, \bibinfo {author} {\bibfnamefont {Duncan}\
  \bibnamefont {England}}, \bibinfo {author} {\bibfnamefont {Robert}\
  \bibnamefont {Fickler}}, \bibinfo {author} {\bibfnamefont {Robert~W}\
  \bibnamefont {Boyd}}, \bibinfo {author} {\bibfnamefont {Berthold-Georg}\
  \bibnamefont {Englert}}, \bibinfo {author} {\bibfnamefont {Luis~L}\
  \bibnamefont {S{\'a}nchez-Soto}}, \ and\ \bibinfo {author} {\bibfnamefont
  {Ebrahim}\ \bibnamefont {Karimi}},\ }\bibfield  {title} {\enquote {\bibinfo
  {title} {Experimental investigation of high-dimensional quantum key
  distribution protocols with twisted photons},}\ }\href@noop {} {\bibfield
  {journal} {\bibinfo  {journal} {Quantum}\ }\textbf {\bibinfo {volume} {2}},\
  \bibinfo {pages} {111} (\bibinfo {year} {2018})}\BibitemShut {NoStop}%
\bibitem [{\citenamefont {Aspuru-Guzik}\ and\ \citenamefont
  {Walther}(2012)}]{aspuru2012photonic}%
  \BibitemOpen
  \bibfield  {author} {\bibinfo {author} {\bibfnamefont {Al{\'a}n}\
  \bibnamefont {Aspuru-Guzik}}\ and\ \bibinfo {author} {\bibfnamefont {Philip}\
  \bibnamefont {Walther}},\ }\bibfield  {title} {\enquote {\bibinfo {title}
  {Photonic quantum simulators},}\ }\href@noop {} {\bibfield  {journal}
  {\bibinfo  {journal} {Nature physics}\ }\textbf {\bibinfo {volume} {8}},\
  \bibinfo {pages} {285--291} (\bibinfo {year} {2012})}\BibitemShut {NoStop}%
\bibitem [{\citenamefont {D'Errico}\ and\ \citenamefont
  {Karimi}(2021)}]{d2021quantum}%
  \BibitemOpen
  \bibfield  {author} {\bibinfo {author} {\bibfnamefont {Alessio}\ \bibnamefont
  {D'Errico}}\ and\ \bibinfo {author} {\bibfnamefont {Ebrahim}\ \bibnamefont
  {Karimi}},\ }\bibfield  {title} {\enquote {\bibinfo {title} {Quantum
  applications of structured photons},}\ }\href@noop {} {\bibfield  {journal}
  {\bibinfo  {journal} {Electromagnetic Vortices: Wave Phenomena and
  Engineering Applications}\ ,\ \bibinfo {pages} {423--455}} (\bibinfo {year}
  {2021})}\BibitemShut {NoStop}%
\bibitem [{\citenamefont {Erhard}\ \emph {et~al.}(2020)\citenamefont {Erhard},
  \citenamefont {Krenn},\ and\ \citenamefont {Zeilinger}}]{erhard2020advances}%
  \BibitemOpen
  \bibfield  {author} {\bibinfo {author} {\bibfnamefont {Manuel}\ \bibnamefont
  {Erhard}}, \bibinfo {author} {\bibfnamefont {Mario}\ \bibnamefont {Krenn}}, \
  and\ \bibinfo {author} {\bibfnamefont {Anton}\ \bibnamefont {Zeilinger}},\
  }\bibfield  {title} {\enquote {\bibinfo {title} {Advances in high-dimensional
  quantum entanglement},}\ }\href@noop {} {\bibfield  {journal} {\bibinfo
  {journal} {Nature Reviews Physics}\ }\textbf {\bibinfo {volume} {2}},\
  \bibinfo {pages} {365--381} (\bibinfo {year} {2020})}\BibitemShut {NoStop}%
\bibitem [{\citenamefont {Hochrainer}\ \emph {et~al.}(2022)\citenamefont
  {Hochrainer}, \citenamefont {Lahiri}, \citenamefont {Erhard}, \citenamefont
  {Krenn},\ and\ \citenamefont {Zeilinger}}]{undetected2022}%
  \BibitemOpen
  \bibfield  {author} {\bibinfo {author} {\bibfnamefont {Armin}\ \bibnamefont
  {Hochrainer}}, \bibinfo {author} {\bibfnamefont {Mayukh}\ \bibnamefont
  {Lahiri}}, \bibinfo {author} {\bibfnamefont {Manuel}\ \bibnamefont {Erhard}},
  \bibinfo {author} {\bibfnamefont {Mario}\ \bibnamefont {Krenn}}, \ and\
  \bibinfo {author} {\bibfnamefont {Anton}\ \bibnamefont {Zeilinger}},\
  }\bibfield  {title} {\enquote {\bibinfo {title} {Quantum indistinguishability
  by path identity and with undetected photons},}\ }\href@noop {} {\bibfield
  {journal} {\bibinfo  {journal} {Reviews of Modern Physics}\ }\textbf
  {\bibinfo {volume} {94}},\ \bibinfo {pages} {025007} (\bibinfo {year}
  {2022})}\BibitemShut {NoStop}%
\bibitem [{\citenamefont {Polino}\ \emph {et~al.}(2020)\citenamefont {Polino},
  \citenamefont {Valeri}, \citenamefont {Spagnolo},\ and\ \citenamefont
  {Sciarrino}}]{polino2020photonic}%
  \BibitemOpen
  \bibfield  {author} {\bibinfo {author} {\bibfnamefont {Emanuele}\
  \bibnamefont {Polino}}, \bibinfo {author} {\bibfnamefont {Mauro}\
  \bibnamefont {Valeri}}, \bibinfo {author} {\bibfnamefont {Nicolò}\
  \bibnamefont {Spagnolo}}, \ and\ \bibinfo {author} {\bibfnamefont {Fabio}\
  \bibnamefont {Sciarrino}},\ }\bibfield  {title} {\enquote {\bibinfo {title}
  {Photonic quantum metrology},}\ }\href {\doibase 10.1116/5.0007577}
  {\bibfield  {journal} {\bibinfo  {journal} {AVS Quantum Science}\ }\textbf
  {\bibinfo {volume} {2}},\ \bibinfo {pages} {024703} (\bibinfo {year}
  {2020})}\BibitemShut {NoStop}%
\bibitem [{\citenamefont {Flamini}\ \emph {et~al.}(2018)\citenamefont
  {Flamini}, \citenamefont {Spagnolo},\ and\ \citenamefont
  {Sciarrino}}]{flamini2018photonic}%
  \BibitemOpen
  \bibfield  {author} {\bibinfo {author} {\bibfnamefont {Fulvio}\ \bibnamefont
  {Flamini}}, \bibinfo {author} {\bibfnamefont {Nicolo}\ \bibnamefont
  {Spagnolo}}, \ and\ \bibinfo {author} {\bibfnamefont {Fabio}\ \bibnamefont
  {Sciarrino}},\ }\bibfield  {title} {\enquote {\bibinfo {title} {Photonic
  quantum information processing: a review},}\ }\href@noop {} {\bibfield
  {journal} {\bibinfo  {journal} {Reports on Progress in Physics}\ }\textbf
  {\bibinfo {volume} {82}},\ \bibinfo {pages} {016001} (\bibinfo {year}
  {2018})}\BibitemShut {NoStop}%
\bibitem [{\citenamefont {Ansari}\ \emph {et~al.}(2018)\citenamefont {Ansari},
  \citenamefont {Donohue}, \citenamefont {Brecht},\ and\ \citenamefont
  {Silberhorn}}]{ansari2018tailoring}%
  \BibitemOpen
  \bibfield  {author} {\bibinfo {author} {\bibfnamefont {Vahid}\ \bibnamefont
  {Ansari}}, \bibinfo {author} {\bibfnamefont {John~M}\ \bibnamefont
  {Donohue}}, \bibinfo {author} {\bibfnamefont {Benjamin}\ \bibnamefont
  {Brecht}}, \ and\ \bibinfo {author} {\bibfnamefont {Christine}\ \bibnamefont
  {Silberhorn}},\ }\bibfield  {title} {\enquote {\bibinfo {title} {Tailoring
  nonlinear processes for quantum optics with pulsed temporal-mode
  encodings},}\ }\href@noop {} {\bibfield  {journal} {\bibinfo  {journal}
  {Optica}\ }\textbf {\bibinfo {volume} {5}},\ \bibinfo {pages} {534--550}
  (\bibinfo {year} {2018})}\BibitemShut {NoStop}%
\bibitem [{\citenamefont {Allen}\ \emph {et~al.}(1992)\citenamefont {Allen},
  \citenamefont {Beijersbergen}, \citenamefont {Spreeuw},\ and\ \citenamefont
  {Woerdman}}]{allen_0AM_1992}%
  \BibitemOpen
  \bibfield  {author} {\bibinfo {author} {\bibfnamefont {L.}~\bibnamefont
  {Allen}}, \bibinfo {author} {\bibfnamefont {M.~W.}\ \bibnamefont
  {Beijersbergen}}, \bibinfo {author} {\bibfnamefont {R.~J.~C.}\ \bibnamefont
  {Spreeuw}}, \ and\ \bibinfo {author} {\bibfnamefont {J.~P.}\ \bibnamefont
  {Woerdman}},\ }\bibfield  {title} {\enquote {\bibinfo {title} {Orbital
  angular momentum of light and the transformation of laguerre-gaussian laser
  modes},}\ }\href {\doibase 10.1103/PhysRevA.45.8185} {\bibfield  {journal}
  {\bibinfo  {journal} {Phys. Rev. A}\ }\textbf {\bibinfo {volume} {45}},\
  \bibinfo {pages} {8185--8189} (\bibinfo {year} {1992})}\BibitemShut {NoStop}%
\bibitem [{\citenamefont {Erhard}\ \emph {et~al.}(2018)\citenamefont {Erhard},
  \citenamefont {Fickler}, \citenamefont {Krenn},\ and\ \citenamefont
  {Zeilinger}}]{erhard2018twisted}%
  \BibitemOpen
  \bibfield  {author} {\bibinfo {author} {\bibfnamefont {Manuel}\ \bibnamefont
  {Erhard}}, \bibinfo {author} {\bibfnamefont {Robert}\ \bibnamefont
  {Fickler}}, \bibinfo {author} {\bibfnamefont {Mario}\ \bibnamefont {Krenn}},
  \ and\ \bibinfo {author} {\bibfnamefont {Anton}\ \bibnamefont {Zeilinger}},\
  }\bibfield  {title} {\enquote {\bibinfo {title} {Twisted photons: new quantum
  perspectives in high dimensions},}\ }\href {\doibase 10.1038/lsa.2017.146}
  {\bibfield  {journal} {\bibinfo  {journal} {Light: Science \& Applications}\
  }\textbf {\bibinfo {volume} {7}},\ \bibinfo {pages} {17146} (\bibinfo {year}
  {2018})}\BibitemShut {NoStop}%
\bibitem [{\citenamefont {Thew}\ \emph {et~al.}(2002)\citenamefont {Thew},
  \citenamefont {Nemoto}, \citenamefont {White},\ and\ \citenamefont
  {Munro}}]{thew2002qudit}%
  \BibitemOpen
  \bibfield  {author} {\bibinfo {author} {\bibfnamefont {RT}~\bibnamefont
  {Thew}}, \bibinfo {author} {\bibfnamefont {Kae}\ \bibnamefont {Nemoto}},
  \bibinfo {author} {\bibfnamefont {Andrew~G}\ \bibnamefont {White}}, \ and\
  \bibinfo {author} {\bibfnamefont {William~J}\ \bibnamefont {Munro}},\
  }\bibfield  {title} {\enquote {\bibinfo {title} {Qudit quantum-state
  tomography},}\ }\href {\doibase https://doi.org/10.1103/PhysRevA.66.012303}
  {\bibfield  {journal} {\bibinfo  {journal} {Physical Review A}\ }\textbf
  {\bibinfo {volume} {66}},\ \bibinfo {pages} {012303} (\bibinfo {year}
  {2002})}\BibitemShut {NoStop}%
\bibitem [{\citenamefont {Nielsen}\ and\ \citenamefont
  {Chuang}(2002)}]{nielsen2002quantum}%
  \BibitemOpen
  \bibfield  {author} {\bibinfo {author} {\bibfnamefont {Michael~A}\
  \bibnamefont {Nielsen}}\ and\ \bibinfo {author} {\bibfnamefont {Isaac}\
  \bibnamefont {Chuang}},\ }\href@noop {} {\enquote {\bibinfo {title} {Quantum
  computation and quantum information},}\ } (\bibinfo {year}
  {2002})\BibitemShut {NoStop}%
\bibitem [{\citenamefont {Eisert}\ \emph {et~al.}(2020)\citenamefont {Eisert},
  \citenamefont {Hangleiter}, \citenamefont {Walk}, \citenamefont {Roth},
  \citenamefont {Markham}, \citenamefont {Parekh}, \citenamefont {Chabaud},\
  and\ \citenamefont {Kashefi}}]{eisert2020quantum}%
  \BibitemOpen
  \bibfield  {author} {\bibinfo {author} {\bibfnamefont {Jens}\ \bibnamefont
  {Eisert}}, \bibinfo {author} {\bibfnamefont {Dominik}\ \bibnamefont
  {Hangleiter}}, \bibinfo {author} {\bibfnamefont {Nathan}\ \bibnamefont
  {Walk}}, \bibinfo {author} {\bibfnamefont {Ingo}\ \bibnamefont {Roth}},
  \bibinfo {author} {\bibfnamefont {Damian}\ \bibnamefont {Markham}}, \bibinfo
  {author} {\bibfnamefont {Rhea}\ \bibnamefont {Parekh}}, \bibinfo {author}
  {\bibfnamefont {Ulysse}\ \bibnamefont {Chabaud}}, \ and\ \bibinfo {author}
  {\bibfnamefont {Elham}\ \bibnamefont {Kashefi}},\ }\bibfield  {title}
  {\enquote {\bibinfo {title} {Quantum certification and benchmarking},}\
  }\href@noop {} {\bibfield  {journal} {\bibinfo  {journal} {Nature Reviews
  Physics}\ }\textbf {\bibinfo {volume} {2}},\ \bibinfo {pages} {382--390}
  (\bibinfo {year} {2020})}\BibitemShut {NoStop}%
\bibitem [{\citenamefont {Agnew}\ \emph {et~al.}(2011)\citenamefont {Agnew},
  \citenamefont {Leach}, \citenamefont {McLaren}, \citenamefont {Roux},\ and\
  \citenamefont {Boyd}}]{agnew2011tomography}%
  \BibitemOpen
  \bibfield  {author} {\bibinfo {author} {\bibfnamefont {Megan}\ \bibnamefont
  {Agnew}}, \bibinfo {author} {\bibfnamefont {Jonathan}\ \bibnamefont {Leach}},
  \bibinfo {author} {\bibfnamefont {Melanie}\ \bibnamefont {McLaren}}, \bibinfo
  {author} {\bibfnamefont {F~Stef}\ \bibnamefont {Roux}}, \ and\ \bibinfo
  {author} {\bibfnamefont {Robert~W}\ \bibnamefont {Boyd}},\ }\bibfield
  {title} {\enquote {\bibinfo {title} {Tomography of the quantum state of
  photons entangled in high dimensions},}\ }\href@noop {} {\bibfield  {journal}
  {\bibinfo  {journal} {Physical Review A}\ }\textbf {\bibinfo {volume} {84}},\
  \bibinfo {pages} {062101} (\bibinfo {year} {2011})}\BibitemShut {NoStop}%
\bibitem [{\citenamefont {Husz{\'a}r}\ and\ \citenamefont
  {Houlsby}(2012)}]{huszar2012adaptive}%
  \BibitemOpen
  \bibfield  {author} {\bibinfo {author} {\bibfnamefont {Ferenc}\ \bibnamefont
  {Husz{\'a}r}}\ and\ \bibinfo {author} {\bibfnamefont {Neil~MT}\ \bibnamefont
  {Houlsby}},\ }\bibfield  {title} {\enquote {\bibinfo {title} {Adaptive
  bayesian quantum tomography},}\ }\href@noop {} {\bibfield  {journal}
  {\bibinfo  {journal} {Physical Review A}\ }\textbf {\bibinfo {volume} {85}},\
  \bibinfo {pages} {052120} (\bibinfo {year} {2012})}\BibitemShut {NoStop}%
\bibitem [{\citenamefont {Mahler}\ \emph {et~al.}(2013)\citenamefont {Mahler},
  \citenamefont {Rozema}, \citenamefont {Darabi}, \citenamefont {Ferrie},
  \citenamefont {Blume-Kohout},\ and\ \citenamefont
  {Steinberg}}]{mahler2013adaptive}%
  \BibitemOpen
  \bibfield  {author} {\bibinfo {author} {\bibfnamefont {Dylan~H}\ \bibnamefont
  {Mahler}}, \bibinfo {author} {\bibfnamefont {Lee~A}\ \bibnamefont {Rozema}},
  \bibinfo {author} {\bibfnamefont {Ardavan}\ \bibnamefont {Darabi}}, \bibinfo
  {author} {\bibfnamefont {Christopher}\ \bibnamefont {Ferrie}}, \bibinfo
  {author} {\bibfnamefont {Robin}\ \bibnamefont {Blume-Kohout}}, \ and\
  \bibinfo {author} {\bibfnamefont {AM}~\bibnamefont {Steinberg}},\ }\bibfield
  {title} {\enquote {\bibinfo {title} {Adaptive quantum state tomography
  improves accuracy quadratically},}\ }\href@noop {} {\bibfield  {journal}
  {\bibinfo  {journal} {Physical review letters}\ }\textbf {\bibinfo {volume}
  {111}},\ \bibinfo {pages} {183601} (\bibinfo {year} {2013})}\BibitemShut
  {NoStop}%
\bibitem [{\citenamefont {Rambach}\ \emph {et~al.}(2021)\citenamefont
  {Rambach}, \citenamefont {Qaryan}, \citenamefont {Kewming}, \citenamefont
  {Ferrie}, \citenamefont {White},\ and\ \citenamefont
  {Romero}}]{rambach2021robust}%
  \BibitemOpen
  \bibfield  {author} {\bibinfo {author} {\bibfnamefont {Markus}\ \bibnamefont
  {Rambach}}, \bibinfo {author} {\bibfnamefont {Mahdi}\ \bibnamefont {Qaryan}},
  \bibinfo {author} {\bibfnamefont {Michael}\ \bibnamefont {Kewming}}, \bibinfo
  {author} {\bibfnamefont {Christopher}\ \bibnamefont {Ferrie}}, \bibinfo
  {author} {\bibfnamefont {Andrew~G}\ \bibnamefont {White}}, \ and\ \bibinfo
  {author} {\bibfnamefont {Jacquiline}\ \bibnamefont {Romero}},\ }\bibfield
  {title} {\enquote {\bibinfo {title} {Robust and efficient high-dimensional
  quantum state tomography},}\ }\href@noop {} {\bibfield  {journal} {\bibinfo
  {journal} {Physical Review Letters}\ }\textbf {\bibinfo {volume} {126}},\
  \bibinfo {pages} {100402} (\bibinfo {year} {2021})}\BibitemShut {NoStop}%
\bibitem [{\citenamefont {Gross}\ \emph {et~al.}(2010)\citenamefont {Gross},
  \citenamefont {Liu}, \citenamefont {Flammia}, \citenamefont {Becker},\ and\
  \citenamefont {Eisert}}]{gross2010quantum}%
  \BibitemOpen
  \bibfield  {author} {\bibinfo {author} {\bibfnamefont {David}\ \bibnamefont
  {Gross}}, \bibinfo {author} {\bibfnamefont {Yi-Kai}\ \bibnamefont {Liu}},
  \bibinfo {author} {\bibfnamefont {Steven~T}\ \bibnamefont {Flammia}},
  \bibinfo {author} {\bibfnamefont {Stephen}\ \bibnamefont {Becker}}, \ and\
  \bibinfo {author} {\bibfnamefont {Jens}\ \bibnamefont {Eisert}},\ }\bibfield
  {title} {\enquote {\bibinfo {title} {Quantum state tomography via compressed
  sensing},}\ }\href@noop {} {\bibfield  {journal} {\bibinfo  {journal}
  {Physical review letters}\ }\textbf {\bibinfo {volume} {105}},\ \bibinfo
  {pages} {150401} (\bibinfo {year} {2010})}\BibitemShut {NoStop}%
\bibitem [{\citenamefont {Bouchard}\ \emph {et~al.}(2019)\citenamefont
  {Bouchard}, \citenamefont {Koutn{\`y}}, \citenamefont {Hufnagel},
  \citenamefont {Hradil}, \citenamefont {{\v{R}}eh{\'a}{\v{c}}ek},
  \citenamefont {Teo}, \citenamefont {Ahn}, \citenamefont {Jeong},
  \citenamefont {S{\'a}nchez-Soto}, \citenamefont {Leuchs} \emph
  {et~al.}}]{bouchard2019compressed}%
  \BibitemOpen
  \bibfield  {author} {\bibinfo {author} {\bibfnamefont {Fr{\'e}d{\'e}ric}\
  \bibnamefont {Bouchard}}, \bibinfo {author} {\bibfnamefont {Dominik}\
  \bibnamefont {Koutn{\`y}}}, \bibinfo {author} {\bibfnamefont {Felix}\
  \bibnamefont {Hufnagel}}, \bibinfo {author} {\bibfnamefont {Zden{\v{e}}k}\
  \bibnamefont {Hradil}}, \bibinfo {author} {\bibfnamefont {Jaroslav}\
  \bibnamefont {{\v{R}}eh{\'a}{\v{c}}ek}}, \bibinfo {author} {\bibfnamefont
  {Yong-Siah}\ \bibnamefont {Teo}}, \bibinfo {author} {\bibfnamefont {Daekun}\
  \bibnamefont {Ahn}}, \bibinfo {author} {\bibfnamefont {Hyunseok}\
  \bibnamefont {Jeong}}, \bibinfo {author} {\bibfnamefont {Luis~L}\
  \bibnamefont {S{\'a}nchez-Soto}}, \bibinfo {author} {\bibfnamefont {Gerd}\
  \bibnamefont {Leuchs}},  \emph {et~al.},\ }\bibfield  {title} {\enquote
  {\bibinfo {title} {Compressed sensing of twisted photons},}\ }\href@noop {}
  {\bibfield  {journal} {\bibinfo  {journal} {Optics Express}\ }\textbf
  {\bibinfo {volume} {27}},\ \bibinfo {pages} {17426--17434} (\bibinfo {year}
  {2019})}\BibitemShut {NoStop}%
\bibitem [{\citenamefont {Bent}\ \emph {et~al.}(2015)\citenamefont {Bent},
  \citenamefont {Qassim}, \citenamefont {Tahir}, \citenamefont {Sych},
  \citenamefont {Leuchs}, \citenamefont {S{\'a}nchez-Soto}, \citenamefont
  {Karimi},\ and\ \citenamefont {Boyd}}]{bent2015experimental}%
  \BibitemOpen
  \bibfield  {author} {\bibinfo {author} {\bibfnamefont {N}~\bibnamefont
  {Bent}}, \bibinfo {author} {\bibfnamefont {H}~\bibnamefont {Qassim}},
  \bibinfo {author} {\bibfnamefont {AA}~\bibnamefont {Tahir}}, \bibinfo
  {author} {\bibfnamefont {D}~\bibnamefont {Sych}}, \bibinfo {author}
  {\bibfnamefont {G}~\bibnamefont {Leuchs}}, \bibinfo {author} {\bibfnamefont
  {Luis~Lorenzo}\ \bibnamefont {S{\'a}nchez-Soto}}, \bibinfo {author}
  {\bibfnamefont {E}~\bibnamefont {Karimi}}, \ and\ \bibinfo {author}
  {\bibfnamefont {RW}~\bibnamefont {Boyd}},\ }\bibfield  {title} {\enquote
  {\bibinfo {title} {Experimental realization of quantum tomography of photonic
  qudits via symmetric informationally complete positive operator-valued
  measures},}\ }\href@noop {} {\bibfield  {journal} {\bibinfo  {journal}
  {Physical Review X}\ }\textbf {\bibinfo {volume} {5}},\ \bibinfo {pages}
  {041006} (\bibinfo {year} {2015})}\BibitemShut {NoStop}%
\bibitem [{\citenamefont {Leith}\ and\ \citenamefont
  {Upatnieks}(1964)}]{leith1964wavefront}%
  \BibitemOpen
  \bibfield  {author} {\bibinfo {author} {\bibfnamefont {Emmett~N}\
  \bibnamefont {Leith}}\ and\ \bibinfo {author} {\bibfnamefont {Juris}\
  \bibnamefont {Upatnieks}},\ }\bibfield  {title} {\enquote {\bibinfo {title}
  {Wavefront reconstruction with diffused illumination and three-dimensional
  objects},}\ }\href@noop {} {\bibfield  {journal} {\bibinfo  {journal} {Josa}\
  }\textbf {\bibinfo {volume} {54}},\ \bibinfo {pages} {1295--1301} (\bibinfo
  {year} {1964})}\BibitemShut {NoStop}%
\bibitem [{\citenamefont {Yamaguchi}(2006)}]{yamaguchi2006phase}%
  \BibitemOpen
  \bibfield  {author} {\bibinfo {author} {\bibfnamefont {Ichirou}\ \bibnamefont
  {Yamaguchi}},\ }\bibfield  {title} {\enquote {\bibinfo {title}
  {Phase-shifting digital holography},}\ }in\ \href@noop {} {\emph {\bibinfo
  {booktitle} {Digital Holography and Three-Dimensional Display}}}\ (\bibinfo
  {publisher} {Springer},\ \bibinfo {year} {2006})\ pp.\ \bibinfo {pages}
  {145--171}\BibitemShut {NoStop}%
\bibitem [{\citenamefont {Verrier}\ and\ \citenamefont
  {Atlan}(2011)}]{verrier2011off}%
  \BibitemOpen
  \bibfield  {author} {\bibinfo {author} {\bibfnamefont {Nicolas}\ \bibnamefont
  {Verrier}}\ and\ \bibinfo {author} {\bibfnamefont {Michael}\ \bibnamefont
  {Atlan}},\ }\bibfield  {title} {\enquote {\bibinfo {title} {Off-axis digital
  hologram reconstruction: some practical considerations},}\ }\href@noop {}
  {\bibfield  {journal} {\bibinfo  {journal} {Applied optics}\ }\textbf
  {\bibinfo {volume} {50}},\ \bibinfo {pages} {H136--H146} (\bibinfo {year}
  {2011})}\BibitemShut {NoStop}%
\bibitem [{\citenamefont {D’Errico}\ \emph {et~al.}(2017)\citenamefont
  {D’Errico}, \citenamefont {D’Amelio}, \citenamefont {Piccirillo},
  \citenamefont {Cardano},\ and\ \citenamefont {Marrucci}}]{d2017measuring}%
  \BibitemOpen
  \bibfield  {author} {\bibinfo {author} {\bibfnamefont {Alessio}\ \bibnamefont
  {D’Errico}}, \bibinfo {author} {\bibfnamefont {Raffaele}\ \bibnamefont
  {D’Amelio}}, \bibinfo {author} {\bibfnamefont {Bruno}\ \bibnamefont
  {Piccirillo}}, \bibinfo {author} {\bibfnamefont {Filippo}\ \bibnamefont
  {Cardano}}, \ and\ \bibinfo {author} {\bibfnamefont {Lorenzo}\ \bibnamefont
  {Marrucci}},\ }\bibfield  {title} {\enquote {\bibinfo {title} {Measuring the
  complex orbital angular momentum spectrum and spatial mode decomposition of
  structured light beams},}\ }\href@noop {} {\bibfield  {journal} {\bibinfo
  {journal} {Optica}\ }\textbf {\bibinfo {volume} {4}},\ \bibinfo {pages}
  {1350--1357} (\bibinfo {year} {2017})}\BibitemShut {NoStop}%
\bibitem [{\citenamefont {Fu}\ \emph {et~al.}(2020)\citenamefont {Fu},
  \citenamefont {Zhai}, \citenamefont {Zhang}, \citenamefont {Liu},
  \citenamefont {Song}, \citenamefont {Zhou},\ and\ \citenamefont
  {Gao}}]{fu2020universal}%
  \BibitemOpen
  \bibfield  {author} {\bibinfo {author} {\bibfnamefont {Shiyao}\ \bibnamefont
  {Fu}}, \bibinfo {author} {\bibfnamefont {Yanwang}\ \bibnamefont {Zhai}},
  \bibinfo {author} {\bibfnamefont {Jianqiang}\ \bibnamefont {Zhang}}, \bibinfo
  {author} {\bibfnamefont {Xueting}\ \bibnamefont {Liu}}, \bibinfo {author}
  {\bibfnamefont {Rui}\ \bibnamefont {Song}}, \bibinfo {author} {\bibfnamefont
  {Heng}\ \bibnamefont {Zhou}}, \ and\ \bibinfo {author} {\bibfnamefont
  {Chunqing}\ \bibnamefont {Gao}},\ }\bibfield  {title} {\enquote {\bibinfo
  {title} {Universal orbital angular momentum spectrum analyzer for beams},}\
  }\href@noop {} {\bibfield  {journal} {\bibinfo  {journal} {PhotoniX}\
  }\textbf {\bibinfo {volume} {1}},\ \bibinfo {pages} {1--12} (\bibinfo {year}
  {2020})}\BibitemShut {NoStop}%
\bibitem [{\citenamefont {Ariyawansa}\ \emph {et~al.}(2021)\citenamefont
  {Ariyawansa}, \citenamefont {Figueroa},\ and\ \citenamefont
  {Brown}}]{ariyawansa2021amplitude}%
  \BibitemOpen
  \bibfield  {author} {\bibinfo {author} {\bibfnamefont {Ashan}\ \bibnamefont
  {Ariyawansa}}, \bibinfo {author} {\bibfnamefont {Edward~J}\ \bibnamefont
  {Figueroa}}, \ and\ \bibinfo {author} {\bibfnamefont {Thomas~G}\ \bibnamefont
  {Brown}},\ }\bibfield  {title} {\enquote {\bibinfo {title} {Amplitude and
  phase sorting of orbital angular momentum states at low light levels},}\
  }\href@noop {} {\bibfield  {journal} {\bibinfo  {journal} {Optica}\ }\textbf
  {\bibinfo {volume} {8}},\ \bibinfo {pages} {147--155} (\bibinfo {year}
  {2021})}\BibitemShut {NoStop}%
\bibitem [{\citenamefont {Brida}\ \emph {et~al.}(2010)\citenamefont {Brida},
  \citenamefont {Genovese},\ and\ \citenamefont
  {Ruo~Berchera}}]{brida2010experimental}%
  \BibitemOpen
  \bibfield  {author} {\bibinfo {author} {\bibfnamefont {Giorgio}\ \bibnamefont
  {Brida}}, \bibinfo {author} {\bibfnamefont {Marco}\ \bibnamefont {Genovese}},
  \ and\ \bibinfo {author} {\bibfnamefont {Ivano}\ \bibnamefont
  {Ruo~Berchera}},\ }\bibfield  {title} {\enquote {\bibinfo {title}
  {Experimental realization of sub-shot-noise quantum imaging},}\ }\href@noop
  {} {\bibfield  {journal} {\bibinfo  {journal} {Nature Photonics}\ }\textbf
  {\bibinfo {volume} {4}},\ \bibinfo {pages} {227--230} (\bibinfo {year}
  {2010})}\BibitemShut {NoStop}%
\bibitem [{\citenamefont {Bolduc}\ \emph {et~al.}(2017)\citenamefont {Bolduc},
  \citenamefont {Faccio},\ and\ \citenamefont {Leach}}]{bolduc2017acquisition}%
  \BibitemOpen
  \bibfield  {author} {\bibinfo {author} {\bibfnamefont {Eliot}\ \bibnamefont
  {Bolduc}}, \bibinfo {author} {\bibfnamefont {Daniele}\ \bibnamefont
  {Faccio}}, \ and\ \bibinfo {author} {\bibfnamefont {Jonathan}\ \bibnamefont
  {Leach}},\ }\bibfield  {title} {\enquote {\bibinfo {title} {Acquisition of
  multiple photon pairs with an emccd camera},}\ }\href@noop {} {\bibfield
  {journal} {\bibinfo  {journal} {Journal of Optics}\ }\textbf {\bibinfo
  {volume} {19}},\ \bibinfo {pages} {054006} (\bibinfo {year}
  {2017})}\BibitemShut {NoStop}%
\bibitem [{\citenamefont {Untern{\"a}hrer}\ \emph {et~al.}(2018)\citenamefont
  {Untern{\"a}hrer}, \citenamefont {Bessire}, \citenamefont {Gasparini},
  \citenamefont {Perenzoni},\ and\ \citenamefont
  {Stefanov}}]{unternahrer2018super}%
  \BibitemOpen
  \bibfield  {author} {\bibinfo {author} {\bibfnamefont {Manuel}\ \bibnamefont
  {Untern{\"a}hrer}}, \bibinfo {author} {\bibfnamefont {B{\"a}nz}\ \bibnamefont
  {Bessire}}, \bibinfo {author} {\bibfnamefont {Leonardo}\ \bibnamefont
  {Gasparini}}, \bibinfo {author} {\bibfnamefont {Matteo}\ \bibnamefont
  {Perenzoni}}, \ and\ \bibinfo {author} {\bibfnamefont {Andr{\'e}}\
  \bibnamefont {Stefanov}},\ }\bibfield  {title} {\enquote {\bibinfo {title}
  {Super-resolution quantum imaging at the heisenberg limit},}\ }\href@noop {}
  {\bibfield  {journal} {\bibinfo  {journal} {Optica}\ }\textbf {\bibinfo
  {volume} {5}},\ \bibinfo {pages} {1150--1154} (\bibinfo {year}
  {2018})}\BibitemShut {NoStop}%
\bibitem [{\citenamefont {Zarghami}\ \emph {et~al.}(2020)\citenamefont
  {Zarghami}, \citenamefont {Gasparini}, \citenamefont {Parmesan},
  \citenamefont {Moreno-Garcia}, \citenamefont {Stefanov}, \citenamefont
  {Bessire}, \citenamefont {Untern{\"a}hrer},\ and\ \citenamefont
  {Perenzoni}}]{zarghami202032}%
  \BibitemOpen
  \bibfield  {author} {\bibinfo {author} {\bibfnamefont {Majid}\ \bibnamefont
  {Zarghami}}, \bibinfo {author} {\bibfnamefont {Leonardo}\ \bibnamefont
  {Gasparini}}, \bibinfo {author} {\bibfnamefont {Luca}\ \bibnamefont
  {Parmesan}}, \bibinfo {author} {\bibfnamefont {Manuel}\ \bibnamefont
  {Moreno-Garcia}}, \bibinfo {author} {\bibfnamefont {Andre}\ \bibnamefont
  {Stefanov}}, \bibinfo {author} {\bibfnamefont {B{\"a}nz}\ \bibnamefont
  {Bessire}}, \bibinfo {author} {\bibfnamefont {Manuel}\ \bibnamefont
  {Untern{\"a}hrer}}, \ and\ \bibinfo {author} {\bibfnamefont {Matteo}\
  \bibnamefont {Perenzoni}},\ }\bibfield  {title} {\enquote {\bibinfo {title}
  {A 32$\times$ 32-pixel cmos imager for quantum optics with per-spad tdc,
  19.48\% fill-factor in a 44.64-$\mu$m pitch reaching 1-mhz observation
  rate},}\ }\href@noop {} {\bibfield  {journal} {\bibinfo  {journal} {IEEE
  Journal of Solid-State Circuits}\ }\textbf {\bibinfo {volume} {55}},\
  \bibinfo {pages} {2819--2830} (\bibinfo {year} {2020})}\BibitemShut {NoStop}%
\bibitem [{\citenamefont {Eckmann}\ \emph {et~al.}(2020)\citenamefont
  {Eckmann}, \citenamefont {Bessire}, \citenamefont {Untern{\"a}hrer},
  \citenamefont {Gasparini}, \citenamefont {Perenzoni},\ and\ \citenamefont
  {Stefanov}}]{eckmann2020characterization}%
  \BibitemOpen
  \bibfield  {author} {\bibinfo {author} {\bibfnamefont {Bruno}\ \bibnamefont
  {Eckmann}}, \bibinfo {author} {\bibfnamefont {B{\"a}nz}\ \bibnamefont
  {Bessire}}, \bibinfo {author} {\bibfnamefont {Manuel}\ \bibnamefont
  {Untern{\"a}hrer}}, \bibinfo {author} {\bibfnamefont {Leonardo}\ \bibnamefont
  {Gasparini}}, \bibinfo {author} {\bibfnamefont {Matteo}\ \bibnamefont
  {Perenzoni}}, \ and\ \bibinfo {author} {\bibfnamefont {Andr{\'e}}\
  \bibnamefont {Stefanov}},\ }\bibfield  {title} {\enquote {\bibinfo {title}
  {Characterization of space-momentum entangled photons with a time resolving
  cmos spad array},}\ }\href@noop {} {\bibfield  {journal} {\bibinfo  {journal}
  {Optics express}\ }\textbf {\bibinfo {volume} {28}},\ \bibinfo {pages}
  {31553--31571} (\bibinfo {year} {2020})}\BibitemShut {NoStop}%
\bibitem [{\citenamefont {Fisher-Levine}\ and\ \citenamefont
  {Nomerotski}(2016)}]{fisher2016timepixcam}%
  \BibitemOpen
  \bibfield  {author} {\bibinfo {author} {\bibfnamefont {M}~\bibnamefont
  {Fisher-Levine}}\ and\ \bibinfo {author} {\bibfnamefont {Andrei}\
  \bibnamefont {Nomerotski}},\ }\bibfield  {title} {\enquote {\bibinfo {title}
  {Timepixcam: a fast optical imager with time-stamping},}\ }\href@noop {}
  {\bibfield  {journal} {\bibinfo  {journal} {Journal of Instrumentation}\
  }\textbf {\bibinfo {volume} {11}},\ \bibinfo {pages} {C03016} (\bibinfo
  {year} {2016})}\BibitemShut {NoStop}%
\bibitem [{\citenamefont {Nomerotski}(2019)}]{nomerotski2019imaging}%
  \BibitemOpen
  \bibfield  {author} {\bibinfo {author} {\bibfnamefont {Andrei}\ \bibnamefont
  {Nomerotski}},\ }\bibfield  {title} {\enquote {\bibinfo {title} {Imaging and
  time stamping of photons with nanosecond resolution in timepix based optical
  cameras},}\ }\href@noop {} {\bibfield  {journal} {\bibinfo  {journal}
  {Nuclear Instruments and Methods in Physics Research Section A: Accelerators,
  Spectrometers, Detectors and Associated Equipment}\ }\textbf {\bibinfo
  {volume} {937}},\ \bibinfo {pages} {26--30} (\bibinfo {year}
  {2019})}\BibitemShut {NoStop}%
\bibitem [{\citenamefont {Nomerotski}\ \emph {et~al.}(2023)\citenamefont
  {Nomerotski}, \citenamefont {Chekhlov}, \citenamefont {Dolzhenko},
  \citenamefont {Glazenborg}, \citenamefont {Farella}, \citenamefont {Keach},
  \citenamefont {Mahon}, \citenamefont {Orlov},\ and\ \citenamefont
  {Svihra}}]{nomerotski2023intensified}%
  \BibitemOpen
  \bibfield  {author} {\bibinfo {author} {\bibfnamefont {Andrei}\ \bibnamefont
  {Nomerotski}}, \bibinfo {author} {\bibfnamefont {Matthew}\ \bibnamefont
  {Chekhlov}}, \bibinfo {author} {\bibfnamefont {Denis}\ \bibnamefont
  {Dolzhenko}}, \bibinfo {author} {\bibfnamefont {Rene}\ \bibnamefont
  {Glazenborg}}, \bibinfo {author} {\bibfnamefont {Brianna}\ \bibnamefont
  {Farella}}, \bibinfo {author} {\bibfnamefont {Michael}\ \bibnamefont
  {Keach}}, \bibinfo {author} {\bibfnamefont {Ryan}\ \bibnamefont {Mahon}},
  \bibinfo {author} {\bibfnamefont {Dmitry}\ \bibnamefont {Orlov}}, \ and\
  \bibinfo {author} {\bibfnamefont {Peter}\ \bibnamefont {Svihra}},\ }\bibfield
   {title} {\enquote {\bibinfo {title} {Intensified tpx3cam, a fast data-driven
  optical camera with nanosecond timing resolution for single photon detection
  in quantum applications},}\ }\href@noop {} {\bibfield  {journal} {\bibinfo
  {journal} {Journal of Instrumentation}\ }\textbf {\bibinfo {volume} {18}},\
  \bibinfo {pages} {C01023} (\bibinfo {year} {2023})}\BibitemShut {NoStop}%
\bibitem [{\citenamefont {Moreau}\ \emph
  {et~al.}(2019{\natexlab{a}})\citenamefont {Moreau}, \citenamefont
  {Toninelli}, \citenamefont {Gregory},\ and\ \citenamefont
  {Padgett}}]{moreau2019imaging}%
  \BibitemOpen
  \bibfield  {author} {\bibinfo {author} {\bibfnamefont {Paul-Antoine}\
  \bibnamefont {Moreau}}, \bibinfo {author} {\bibfnamefont {Ermes}\
  \bibnamefont {Toninelli}}, \bibinfo {author} {\bibfnamefont {Thomas}\
  \bibnamefont {Gregory}}, \ and\ \bibinfo {author} {\bibfnamefont {Miles~J}\
  \bibnamefont {Padgett}},\ }\bibfield  {title} {\enquote {\bibinfo {title}
  {Imaging with quantum states of light},}\ }\href@noop {} {\bibfield
  {journal} {\bibinfo  {journal} {Nature Reviews Physics}\ }\textbf {\bibinfo
  {volume} {1}},\ \bibinfo {pages} {367--380} (\bibinfo {year}
  {2019}{\natexlab{a}})}\BibitemShut {NoStop}%
\bibitem [{\citenamefont {Zhang}\ \emph {et~al.}(2020)\citenamefont {Zhang},
  \citenamefont {England}, \citenamefont {Nomerotski}, \citenamefont {Svihra},
  \citenamefont {Ferrante}, \citenamefont {Hockett},\ and\ \citenamefont
  {Sussman}}]{zhang2020multidimensional}%
  \BibitemOpen
  \bibfield  {author} {\bibinfo {author} {\bibfnamefont {Yingwen}\ \bibnamefont
  {Zhang}}, \bibinfo {author} {\bibfnamefont {Duncan}\ \bibnamefont {England}},
  \bibinfo {author} {\bibfnamefont {Andrei}\ \bibnamefont {Nomerotski}},
  \bibinfo {author} {\bibfnamefont {Peter}\ \bibnamefont {Svihra}}, \bibinfo
  {author} {\bibfnamefont {Steven}\ \bibnamefont {Ferrante}}, \bibinfo {author}
  {\bibfnamefont {Paul}\ \bibnamefont {Hockett}}, \ and\ \bibinfo {author}
  {\bibfnamefont {Benjamin}\ \bibnamefont {Sussman}},\ }\bibfield  {title}
  {\enquote {\bibinfo {title} {Multidimensional quantum-enhanced target
  detection via spectrotemporal-correlation measurements},}\ }\href@noop {}
  {\bibfield  {journal} {\bibinfo  {journal} {Physical Review A}\ }\textbf
  {\bibinfo {volume} {101}},\ \bibinfo {pages} {053808} (\bibinfo {year}
  {2020})}\BibitemShut {NoStop}%
\bibitem [{\citenamefont {Defienne}\ \emph {et~al.}(2019)\citenamefont
  {Defienne}, \citenamefont {Reichert}, \citenamefont {Fleischer},\ and\
  \citenamefont {Faccio}}]{defienne2019quantum}%
  \BibitemOpen
  \bibfield  {author} {\bibinfo {author} {\bibfnamefont {Hugo}\ \bibnamefont
  {Defienne}}, \bibinfo {author} {\bibfnamefont {Matthew}\ \bibnamefont
  {Reichert}}, \bibinfo {author} {\bibfnamefont {Jason~W}\ \bibnamefont
  {Fleischer}}, \ and\ \bibinfo {author} {\bibfnamefont {Daniele}\ \bibnamefont
  {Faccio}},\ }\bibfield  {title} {\enquote {\bibinfo {title} {Quantum image
  distillation},}\ }\href@noop {} {\bibfield  {journal} {\bibinfo  {journal}
  {Science advances}\ }\textbf {\bibinfo {volume} {5}},\ \bibinfo {pages}
  {eaax0307} (\bibinfo {year} {2019})}\BibitemShut {NoStop}%
\bibitem [{\citenamefont {Salari}\ \emph {et~al.}(2021)\citenamefont {Salari},
  \citenamefont {Paneru}, \citenamefont {Saglamyurek}, \citenamefont {Ghadimi},
  \citenamefont {Abdar}, \citenamefont {Rezaee}, \citenamefont {Aslani},
  \citenamefont {Barzanjeh},\ and\ \citenamefont {Karimi}}]{salari2021quantum}%
  \BibitemOpen
  \bibfield  {author} {\bibinfo {author} {\bibfnamefont {Vahid}\ \bibnamefont
  {Salari}}, \bibinfo {author} {\bibfnamefont {Dilip}\ \bibnamefont {Paneru}},
  \bibinfo {author} {\bibfnamefont {Erhan}\ \bibnamefont {Saglamyurek}},
  \bibinfo {author} {\bibfnamefont {Milad}\ \bibnamefont {Ghadimi}}, \bibinfo
  {author} {\bibfnamefont {Moloud}\ \bibnamefont {Abdar}}, \bibinfo {author}
  {\bibfnamefont {Mohammadreza}\ \bibnamefont {Rezaee}}, \bibinfo {author}
  {\bibfnamefont {Mehdi}\ \bibnamefont {Aslani}}, \bibinfo {author}
  {\bibfnamefont {Shabir}\ \bibnamefont {Barzanjeh}}, \ and\ \bibinfo {author}
  {\bibfnamefont {Ebrahim}\ \bibnamefont {Karimi}},\ }\bibfield  {title}
  {\enquote {\bibinfo {title} {Quantum face recognition protocol with ghost
  imaging},}\ }\href@noop {} {\bibfield  {journal} {\bibinfo  {journal} {arXiv
  preprint arXiv:2110.10088}\ } (\bibinfo {year} {2021})}\BibitemShut {NoStop}%
\bibitem [{\citenamefont {Tenne}\ \emph {et~al.}(2019)\citenamefont {Tenne},
  \citenamefont {Rossman}, \citenamefont {Rephael}, \citenamefont {Israel},
  \citenamefont {Krupinski-Ptaszek}, \citenamefont {Lapkiewicz}, \citenamefont
  {Silberberg},\ and\ \citenamefont {Oron}}]{tenne2019super}%
  \BibitemOpen
  \bibfield  {author} {\bibinfo {author} {\bibfnamefont {Ron}\ \bibnamefont
  {Tenne}}, \bibinfo {author} {\bibfnamefont {Uri}\ \bibnamefont {Rossman}},
  \bibinfo {author} {\bibfnamefont {Batel}\ \bibnamefont {Rephael}}, \bibinfo
  {author} {\bibfnamefont {Yonatan}\ \bibnamefont {Israel}}, \bibinfo {author}
  {\bibfnamefont {Alexander}\ \bibnamefont {Krupinski-Ptaszek}}, \bibinfo
  {author} {\bibfnamefont {Radek}\ \bibnamefont {Lapkiewicz}}, \bibinfo
  {author} {\bibfnamefont {Yaron}\ \bibnamefont {Silberberg}}, \ and\ \bibinfo
  {author} {\bibfnamefont {Dan}\ \bibnamefont {Oron}},\ }\bibfield  {title}
  {\enquote {\bibinfo {title} {Super-resolution enhancement by quantum image
  scanning microscopy},}\ }\href@noop {} {\bibfield  {journal} {\bibinfo
  {journal} {Nature Photonics}\ }\textbf {\bibinfo {volume} {13}},\ \bibinfo
  {pages} {116--122} (\bibinfo {year} {2019})}\BibitemShut {NoStop}%
\bibitem [{\citenamefont {Toninelli}\ \emph {et~al.}(2019)\citenamefont
  {Toninelli}, \citenamefont {Moreau}, \citenamefont {Gregory}, \citenamefont
  {Mihalyi}, \citenamefont {Edgar}, \citenamefont {Radwell},\ and\
  \citenamefont {Padgett}}]{toninelli2019resolution}%
  \BibitemOpen
  \bibfield  {author} {\bibinfo {author} {\bibfnamefont {Ermes}\ \bibnamefont
  {Toninelli}}, \bibinfo {author} {\bibfnamefont {Paul-Antoine}\ \bibnamefont
  {Moreau}}, \bibinfo {author} {\bibfnamefont {Thomas}\ \bibnamefont
  {Gregory}}, \bibinfo {author} {\bibfnamefont {Adam}\ \bibnamefont {Mihalyi}},
  \bibinfo {author} {\bibfnamefont {Matthew}\ \bibnamefont {Edgar}}, \bibinfo
  {author} {\bibfnamefont {Neal}\ \bibnamefont {Radwell}}, \ and\ \bibinfo
  {author} {\bibfnamefont {Miles}\ \bibnamefont {Padgett}},\ }\bibfield
  {title} {\enquote {\bibinfo {title} {Resolution-enhanced quantum imaging by
  centroid estimation of biphotons},}\ }\href@noop {} {\bibfield  {journal}
  {\bibinfo  {journal} {Optica}\ }\textbf {\bibinfo {volume} {6}},\ \bibinfo
  {pages} {347--353} (\bibinfo {year} {2019})}\BibitemShut {NoStop}%
\bibitem [{\citenamefont {Defienne}\ \emph {et~al.}(2022)\citenamefont
  {Defienne}, \citenamefont {Cameron}, \citenamefont {Ndagano}, \citenamefont
  {Lyons}, \citenamefont {Reichert}, \citenamefont {Zhao}, \citenamefont
  {Harvey}, \citenamefont {Charbon}, \citenamefont {Fleischer},\ and\
  \citenamefont {Faccio}}]{defienne2022pixel}%
  \BibitemOpen
  \bibfield  {author} {\bibinfo {author} {\bibfnamefont {Hugo}\ \bibnamefont
  {Defienne}}, \bibinfo {author} {\bibfnamefont {Patrick}\ \bibnamefont
  {Cameron}}, \bibinfo {author} {\bibfnamefont {Bienvenu}\ \bibnamefont
  {Ndagano}}, \bibinfo {author} {\bibfnamefont {Ashley}\ \bibnamefont {Lyons}},
  \bibinfo {author} {\bibfnamefont {Matthew}\ \bibnamefont {Reichert}},
  \bibinfo {author} {\bibfnamefont {Jiuxuan}\ \bibnamefont {Zhao}}, \bibinfo
  {author} {\bibfnamefont {Andrew~R}\ \bibnamefont {Harvey}}, \bibinfo {author}
  {\bibfnamefont {Edoardo}\ \bibnamefont {Charbon}}, \bibinfo {author}
  {\bibfnamefont {Jason~W}\ \bibnamefont {Fleischer}}, \ and\ \bibinfo {author}
  {\bibfnamefont {Daniele}\ \bibnamefont {Faccio}},\ }\bibfield  {title}
  {\enquote {\bibinfo {title} {Pixel super-resolution with spatially entangled
  photons},}\ }\href@noop {} {\bibfield  {journal} {\bibinfo  {journal} {Nature
  communications}\ }\textbf {\bibinfo {volume} {13}},\ \bibinfo {pages} {1--9}
  (\bibinfo {year} {2022})}\BibitemShut {NoStop}%
\bibitem [{\citenamefont {Boucher}\ \emph {et~al.}(2021)\citenamefont
  {Boucher}, \citenamefont {Defienne},\ and\ \citenamefont
  {Gigan}}]{boucher2021engineering}%
  \BibitemOpen
  \bibfield  {author} {\bibinfo {author} {\bibfnamefont {Pauline}\ \bibnamefont
  {Boucher}}, \bibinfo {author} {\bibfnamefont {Hugo}\ \bibnamefont
  {Defienne}}, \ and\ \bibinfo {author} {\bibfnamefont {Sylvain}\ \bibnamefont
  {Gigan}},\ }\bibfield  {title} {\enquote {\bibinfo {title} {Engineering
  spatial correlations of entangled photon pairs by pump beam shaping},}\
  }\href@noop {} {\bibfield  {journal} {\bibinfo  {journal} {Optics Letters}\
  }\textbf {\bibinfo {volume} {46}},\ \bibinfo {pages} {4200--4203} (\bibinfo
  {year} {2021})}\BibitemShut {NoStop}%
\bibitem [{\citenamefont {Devaux}\ \emph {et~al.}(2020)\citenamefont {Devaux},
  \citenamefont {Mosset}, \citenamefont {Moreau},\ and\ \citenamefont
  {Lantz}}]{devaux2020imaging}%
  \BibitemOpen
  \bibfield  {author} {\bibinfo {author} {\bibfnamefont {Fabrice}\ \bibnamefont
  {Devaux}}, \bibinfo {author} {\bibfnamefont {Alexis}\ \bibnamefont {Mosset}},
  \bibinfo {author} {\bibfnamefont {Paul-Antoine}\ \bibnamefont {Moreau}}, \
  and\ \bibinfo {author} {\bibfnamefont {Eric}\ \bibnamefont {Lantz}},\
  }\bibfield  {title} {\enquote {\bibinfo {title} {Imaging spatiotemporal
  hong-ou-mandel interference of biphoton states of extremely high schmidt
  number},}\ }\href@noop {} {\bibfield  {journal} {\bibinfo  {journal}
  {Physical Review X}\ }\textbf {\bibinfo {volume} {10}},\ \bibinfo {pages}
  {031031} (\bibinfo {year} {2020})}\BibitemShut {NoStop}%
\bibitem [{\citenamefont {Zhang}\ \emph {et~al.}(2021)\citenamefont {Zhang},
  \citenamefont {England}, \citenamefont {Nomerotski},\ and\ \citenamefont
  {Sussman}}]{zhang2021high}%
  \BibitemOpen
  \bibfield  {author} {\bibinfo {author} {\bibfnamefont {Yingwen}\ \bibnamefont
  {Zhang}}, \bibinfo {author} {\bibfnamefont {Duncan}\ \bibnamefont {England}},
  \bibinfo {author} {\bibfnamefont {Andrei}\ \bibnamefont {Nomerotski}}, \ and\
  \bibinfo {author} {\bibfnamefont {Benjamin}\ \bibnamefont {Sussman}},\
  }\bibfield  {title} {\enquote {\bibinfo {title} {High speed imaging of
  spectral-temporal correlations in hong-ou-mandel interference},}\ }\href@noop
  {} {\bibfield  {journal} {\bibinfo  {journal} {Optics Express}\ }\textbf
  {\bibinfo {volume} {29}},\ \bibinfo {pages} {28217--28227} (\bibinfo {year}
  {2021})}\BibitemShut {NoStop}%
\bibitem [{\citenamefont {Gao}\ \emph {et~al.}(2022)\citenamefont {Gao},
  \citenamefont {Zhang}, \citenamefont {D’Errico}, \citenamefont {Heshami},\
  and\ \citenamefont {Karimi}}]{gao2022high}%
  \BibitemOpen
  \bibfield  {author} {\bibinfo {author} {\bibfnamefont {Xiaoqin}\ \bibnamefont
  {Gao}}, \bibinfo {author} {\bibfnamefont {Yingwen}\ \bibnamefont {Zhang}},
  \bibinfo {author} {\bibfnamefont {Alessio}\ \bibnamefont {D’Errico}},
  \bibinfo {author} {\bibfnamefont {Khabat}\ \bibnamefont {Heshami}}, \ and\
  \bibinfo {author} {\bibfnamefont {Ebrahim}\ \bibnamefont {Karimi}},\
  }\bibfield  {title} {\enquote {\bibinfo {title} {High-speed imaging of
  spatiotemporal correlations in hong-ou-mandel interference},}\ }\href@noop {}
  {\bibfield  {journal} {\bibinfo  {journal} {Optics Express}\ }\textbf
  {\bibinfo {volume} {30}},\ \bibinfo {pages} {19456--19464} (\bibinfo {year}
  {2022})}\BibitemShut {NoStop}%
\bibitem [{\citenamefont {Zhang}\ \emph {et~al.}(2022)\citenamefont {Zhang},
  \citenamefont {Orth}, \citenamefont {England},\ and\ \citenamefont
  {Sussman}}]{zhang2022ray}%
  \BibitemOpen
  \bibfield  {author} {\bibinfo {author} {\bibfnamefont {Yingwen}\ \bibnamefont
  {Zhang}}, \bibinfo {author} {\bibfnamefont {Antony}\ \bibnamefont {Orth}},
  \bibinfo {author} {\bibfnamefont {Duncan}\ \bibnamefont {England}}, \ and\
  \bibinfo {author} {\bibfnamefont {Benjamin}\ \bibnamefont {Sussman}},\
  }\bibfield  {title} {\enquote {\bibinfo {title} {Ray tracing with quantum
  correlated photons to image a three-dimensional scene},}\ }\href@noop {}
  {\bibfield  {journal} {\bibinfo  {journal} {Physical Review A}\ }\textbf
  {\bibinfo {volume} {105}},\ \bibinfo {pages} {L011701} (\bibinfo {year}
  {2022})}\BibitemShut {NoStop}%
\bibitem [{\citenamefont {Moreau}\ \emph
  {et~al.}(2019{\natexlab{b}})\citenamefont {Moreau}, \citenamefont
  {Toninelli}, \citenamefont {Gregory}, \citenamefont {Aspden}, \citenamefont
  {Morris},\ and\ \citenamefont {Padgett}}]{moreau2019imagingbell}%
  \BibitemOpen
  \bibfield  {author} {\bibinfo {author} {\bibfnamefont {Paul-Antoine}\
  \bibnamefont {Moreau}}, \bibinfo {author} {\bibfnamefont {Ermes}\
  \bibnamefont {Toninelli}}, \bibinfo {author} {\bibfnamefont {Thomas}\
  \bibnamefont {Gregory}}, \bibinfo {author} {\bibfnamefont {Reuben~S}\
  \bibnamefont {Aspden}}, \bibinfo {author} {\bibfnamefont {Peter~A}\
  \bibnamefont {Morris}}, \ and\ \bibinfo {author} {\bibfnamefont {Miles~J}\
  \bibnamefont {Padgett}},\ }\bibfield  {title} {\enquote {\bibinfo {title}
  {Imaging bell-type nonlocal behavior},}\ }\href@noop {} {\bibfield  {journal}
  {\bibinfo  {journal} {Science advances}\ }\textbf {\bibinfo {volume} {5}},\
  \bibinfo {pages} {eaaw2563} (\bibinfo {year}
  {2019}{\natexlab{b}})}\BibitemShut {NoStop}%
\bibitem [{\citenamefont {Defienne}\ \emph {et~al.}(2021)\citenamefont
  {Defienne}, \citenamefont {Ndagano}, \citenamefont {Lyons},\ and\
  \citenamefont {Faccio}}]{defienne2021polarization}%
  \BibitemOpen
  \bibfield  {author} {\bibinfo {author} {\bibfnamefont {Hugo}\ \bibnamefont
  {Defienne}}, \bibinfo {author} {\bibfnamefont {Bienvenu}\ \bibnamefont
  {Ndagano}}, \bibinfo {author} {\bibfnamefont {Ashley}\ \bibnamefont {Lyons}},
  \ and\ \bibinfo {author} {\bibfnamefont {Daniele}\ \bibnamefont {Faccio}},\
  }\bibfield  {title} {\enquote {\bibinfo {title} {Polarization
  entanglement-enabled quantum holography},}\ }\href@noop {} {\bibfield
  {journal} {\bibinfo  {journal} {Nature Physics}\ }\textbf {\bibinfo {volume}
  {17}},\ \bibinfo {pages} {591--597} (\bibinfo {year} {2021})}\BibitemShut
  {NoStop}%
\bibitem [{\citenamefont {T{\"o}pfer}\ \emph {et~al.}(2022)\citenamefont
  {T{\"o}pfer}, \citenamefont {Gilaberte~Basset}, \citenamefont {Fuenzalida},
  \citenamefont {Steinlechner}, \citenamefont {Torres},\ and\ \citenamefont
  {Gr{\"a}fe}}]{topfer2022quantum}%
  \BibitemOpen
  \bibfield  {author} {\bibinfo {author} {\bibfnamefont {Sebastian}\
  \bibnamefont {T{\"o}pfer}}, \bibinfo {author} {\bibfnamefont {Marta}\
  \bibnamefont {Gilaberte~Basset}}, \bibinfo {author} {\bibfnamefont {Jorge}\
  \bibnamefont {Fuenzalida}}, \bibinfo {author} {\bibfnamefont {Fabian}\
  \bibnamefont {Steinlechner}}, \bibinfo {author} {\bibfnamefont {Juan~P}\
  \bibnamefont {Torres}}, \ and\ \bibinfo {author} {\bibfnamefont {Markus}\
  \bibnamefont {Gr{\"a}fe}},\ }\bibfield  {title} {\enquote {\bibinfo {title}
  {Quantum holography with undetected light},}\ }\href@noop {} {\bibfield
  {journal} {\bibinfo  {journal} {Science advances}\ }\textbf {\bibinfo
  {volume} {8}},\ \bibinfo {pages} {eabl4301} (\bibinfo {year}
  {2022})}\BibitemShut {NoStop}%
\bibitem [{\citenamefont {Thekkadath}\ \emph {et~al.}(2023)\citenamefont
  {Thekkadath}, \citenamefont {England}, \citenamefont {Bouchard},
  \citenamefont {Zhang}, \citenamefont {Kim},\ and\ \citenamefont
  {Sussman}}]{thekkadath2023intensity}%
  \BibitemOpen
  \bibfield  {author} {\bibinfo {author} {\bibfnamefont {GS}~\bibnamefont
  {Thekkadath}}, \bibinfo {author} {\bibfnamefont {D}~\bibnamefont {England}},
  \bibinfo {author} {\bibfnamefont {F}~\bibnamefont {Bouchard}}, \bibinfo
  {author} {\bibfnamefont {Y}~\bibnamefont {Zhang}}, \bibinfo {author}
  {\bibfnamefont {MS}~\bibnamefont {Kim}}, \ and\ \bibinfo {author}
  {\bibfnamefont {B}~\bibnamefont {Sussman}},\ }\bibfield  {title} {\enquote
  {\bibinfo {title} {Intensity interferometry for holography with quantum and
  classical light},}\ }\href@noop {} {\bibfield  {journal} {\bibinfo  {journal}
  {arXiv preprint arXiv:2301.10068}\ } (\bibinfo {year} {2023})}\BibitemShut
  {NoStop}%
\bibitem [{\citenamefont {Parniak}\ \emph {et~al.}(2017)\citenamefont
  {Parniak}, \citenamefont {Dabrowski}, \citenamefont {Mazelanik},
  \citenamefont {Leszczy{\'n}ski}, \citenamefont {Lipka},\ and\ \citenamefont
  {Wasilewski}}]{parniak2017wavevector}%
  \BibitemOpen
  \bibfield  {author} {\bibinfo {author} {\bibfnamefont {Michal}\ \bibnamefont
  {Parniak}}, \bibinfo {author} {\bibfnamefont {Michal}\ \bibnamefont
  {Dabrowski}}, \bibinfo {author} {\bibfnamefont {Mateusz}\ \bibnamefont
  {Mazelanik}}, \bibinfo {author} {\bibfnamefont {Adam}\ \bibnamefont
  {Leszczy{\'n}ski}}, \bibinfo {author} {\bibfnamefont {Michal}\ \bibnamefont
  {Lipka}}, \ and\ \bibinfo {author} {\bibfnamefont {Wojciech}\ \bibnamefont
  {Wasilewski}},\ }\bibfield  {title} {\enquote {\bibinfo {title} {Wavevector
  multiplexed atomic quantum memory via spatially-resolved single-photon
  detection},}\ }\href@noop {} {\bibfield  {journal} {\bibinfo  {journal}
  {Nature communications}\ }\textbf {\bibinfo {volume} {8}},\ \bibinfo {pages}
  {1--9} (\bibinfo {year} {2017})}\BibitemShut {NoStop}%
\bibitem [{\citenamefont {Walborn}\ \emph {et~al.}(2010)\citenamefont
  {Walborn}, \citenamefont {Monken}, \citenamefont {P{\'a}dua},\ and\
  \citenamefont {Ribeiro}}]{walborn2010spatial}%
  \BibitemOpen
  \bibfield  {author} {\bibinfo {author} {\bibfnamefont {Stephen~P}\
  \bibnamefont {Walborn}}, \bibinfo {author} {\bibfnamefont {CH}~\bibnamefont
  {Monken}}, \bibinfo {author} {\bibfnamefont {S}~\bibnamefont {P{\'a}dua}}, \
  and\ \bibinfo {author} {\bibfnamefont {PH~Souto}\ \bibnamefont {Ribeiro}},\
  }\bibfield  {title} {\enquote {\bibinfo {title} {Spatial correlations in
  parametric down-conversion},}\ }\href@noop {} {\bibfield  {journal} {\bibinfo
   {journal} {Physics Reports}\ }\textbf {\bibinfo {volume} {495}},\ \bibinfo
  {pages} {87--139} (\bibinfo {year} {2010})}\BibitemShut {NoStop}%
\bibitem [{\citenamefont {Mair}\ \emph {et~al.}(2001)\citenamefont {Mair},
  \citenamefont {Vaziri}, \citenamefont {Weihs},\ and\ \citenamefont
  {Zeilinger}}]{mair2001entanglement}%
  \BibitemOpen
  \bibfield  {author} {\bibinfo {author} {\bibfnamefont {Alois}\ \bibnamefont
  {Mair}}, \bibinfo {author} {\bibfnamefont {Alipasha}\ \bibnamefont {Vaziri}},
  \bibinfo {author} {\bibfnamefont {Gregor}\ \bibnamefont {Weihs}}, \ and\
  \bibinfo {author} {\bibfnamefont {Anton}\ \bibnamefont {Zeilinger}},\
  }\bibfield  {title} {\enquote {\bibinfo {title} {Entanglement of the orbital
  angular momentum states of photons},}\ }\href {\doibase 10.1038/35085529}
  {\bibfield  {journal} {\bibinfo  {journal} {Nature}\ }\textbf {\bibinfo
  {volume} {412}},\ \bibinfo {pages} {313} (\bibinfo {year}
  {2001})}\BibitemShut {NoStop}%
\bibitem [{\citenamefont {D’Errico}\ \emph {et~al.}(2021)\citenamefont
  {D’Errico}, \citenamefont {Hufnagel}, \citenamefont {Miatto}, \citenamefont
  {Rezaee},\ and\ \citenamefont {Karimi}}]{d2021full}%
  \BibitemOpen
  \bibfield  {author} {\bibinfo {author} {\bibfnamefont {Alessio}\ \bibnamefont
  {D’Errico}}, \bibinfo {author} {\bibfnamefont {Felix}\ \bibnamefont
  {Hufnagel}}, \bibinfo {author} {\bibfnamefont {Filippo}\ \bibnamefont
  {Miatto}}, \bibinfo {author} {\bibfnamefont {Mohammadreza}\ \bibnamefont
  {Rezaee}}, \ and\ \bibinfo {author} {\bibfnamefont {Ebrahim}\ \bibnamefont
  {Karimi}},\ }\bibfield  {title} {\enquote {\bibinfo {title} {Full-mode
  characterization of correlated photon pairs generated in spontaneous
  downconversion},}\ }\href@noop {} {\bibfield  {journal} {\bibinfo  {journal}
  {Optics Letters}\ }\textbf {\bibinfo {volume} {46}},\ \bibinfo {pages}
  {2388--2391} (\bibinfo {year} {2021})}\BibitemShut {NoStop}%
\bibitem [{\citenamefont {Zhang}\ \emph {et~al.}(2014)\citenamefont {Zhang},
  \citenamefont {Roux}, \citenamefont {McLaren},\ and\ \citenamefont
  {Forbes}}]{zhang2014radial}%
  \BibitemOpen
  \bibfield  {author} {\bibinfo {author} {\bibfnamefont {Yingwen}\ \bibnamefont
  {Zhang}}, \bibinfo {author} {\bibfnamefont {Filippus~S}\ \bibnamefont
  {Roux}}, \bibinfo {author} {\bibfnamefont {Melanie}\ \bibnamefont {McLaren}},
  \ and\ \bibinfo {author} {\bibfnamefont {Andrew}\ \bibnamefont {Forbes}},\
  }\bibfield  {title} {\enquote {\bibinfo {title} {Radial modal dependence of
  the azimuthal spectrum after parametric down-conversion},}\ }\href@noop {}
  {\bibfield  {journal} {\bibinfo  {journal} {Physical Review A}\ }\textbf
  {\bibinfo {volume} {89}},\ \bibinfo {pages} {043820} (\bibinfo {year}
  {2014})}\BibitemShut {NoStop}%
\bibitem [{\citenamefont {Salakhutdinov}\ \emph {et~al.}(2012)\citenamefont
  {Salakhutdinov}, \citenamefont {Eliel},\ and\ \citenamefont
  {L{\"o}ffler}}]{salakhutdinov2012full}%
  \BibitemOpen
  \bibfield  {author} {\bibinfo {author} {\bibfnamefont {VD}~\bibnamefont
  {Salakhutdinov}}, \bibinfo {author} {\bibfnamefont {ER}~\bibnamefont
  {Eliel}}, \ and\ \bibinfo {author} {\bibfnamefont {W}~\bibnamefont
  {L{\"o}ffler}},\ }\bibfield  {title} {\enquote {\bibinfo {title} {Full-field
  quantum correlations of spatially entangled photons},}\ }\href@noop {}
  {\bibfield  {journal} {\bibinfo  {journal} {Physical review letters}\
  }\textbf {\bibinfo {volume} {108}},\ \bibinfo {pages} {173604} (\bibinfo
  {year} {2012})}\BibitemShut {NoStop}%
\bibitem [{\citenamefont {Karimi}\ \emph {et~al.}(2007)\citenamefont {Karimi},
  \citenamefont {Zito}, \citenamefont {Piccirillo}, \citenamefont {Marrucci},\
  and\ \citenamefont {Santamato}}]{karimi_07}%
  \BibitemOpen
  \bibfield  {author} {\bibinfo {author} {\bibfnamefont {Ebrahim}\ \bibnamefont
  {Karimi}}, \bibinfo {author} {\bibfnamefont {Gianluigi}\ \bibnamefont
  {Zito}}, \bibinfo {author} {\bibfnamefont {Bruno}\ \bibnamefont
  {Piccirillo}}, \bibinfo {author} {\bibfnamefont {Lorenzo}\ \bibnamefont
  {Marrucci}}, \ and\ \bibinfo {author} {\bibfnamefont {Enrico}\ \bibnamefont
  {Santamato}},\ }\bibfield  {title} {\enquote {\bibinfo {title}
  {Hypergeometric-gaussian modes},}\ }\href {\doibase 10.1364/OL.32.003053}
  {\bibfield  {journal} {\bibinfo  {journal} {Opt. Lett.}\ }\textbf {\bibinfo
  {volume} {32}},\ \bibinfo {pages} {3053--3055} (\bibinfo {year}
  {2007})}\BibitemShut {NoStop}%
\bibitem [{\citenamefont {Siegman}(1986)}]{siegman1986lasers}%
  \BibitemOpen
  \bibfield  {author} {\bibinfo {author} {\bibfnamefont {Anthony~E}\
  \bibnamefont {Siegman}},\ }\href@noop {} {\emph {\bibinfo {title} {Lasers}}}\
  (\bibinfo  {publisher} {University science books},\ \bibinfo {year}
  {1986})\BibitemShut {NoStop}%
\bibitem [{\citenamefont {Vallone}\ \emph {et~al.}(2016)\citenamefont
  {Vallone}, \citenamefont {Parisi}, \citenamefont {Spinello}, \citenamefont
  {Mari}, \citenamefont {Tamburini},\ and\ \citenamefont
  {Villoresi}}]{vallone2016general}%
  \BibitemOpen
  \bibfield  {author} {\bibinfo {author} {\bibfnamefont {Giuseppe}\
  \bibnamefont {Vallone}}, \bibinfo {author} {\bibfnamefont {Giuseppe}\
  \bibnamefont {Parisi}}, \bibinfo {author} {\bibfnamefont {Fabio}\
  \bibnamefont {Spinello}}, \bibinfo {author} {\bibfnamefont {Elettra}\
  \bibnamefont {Mari}}, \bibinfo {author} {\bibfnamefont {Fabrizio}\
  \bibnamefont {Tamburini}}, \ and\ \bibinfo {author} {\bibfnamefont {Paolo}\
  \bibnamefont {Villoresi}},\ }\bibfield  {title} {\enquote {\bibinfo {title}
  {General theorem on the divergence of vortex beams},}\ }\href@noop {}
  {\bibfield  {journal} {\bibinfo  {journal} {Physical Review A}\ }\textbf
  {\bibinfo {volume} {94}},\ \bibinfo {pages} {023802} (\bibinfo {year}
  {2016})}\BibitemShut {NoStop}%
\bibitem [{\citenamefont {Dada}\ \emph {et~al.}(2011)\citenamefont {Dada},
  \citenamefont {Leach}, \citenamefont {Buller}, \citenamefont {Padgett},\ and\
  \citenamefont {Andersson}}]{dada2011experimental}%
  \BibitemOpen
  \bibfield  {author} {\bibinfo {author} {\bibfnamefont {Adetunmise~C}\
  \bibnamefont {Dada}}, \bibinfo {author} {\bibfnamefont {Jonathan}\
  \bibnamefont {Leach}}, \bibinfo {author} {\bibfnamefont {Gerald~S}\
  \bibnamefont {Buller}}, \bibinfo {author} {\bibfnamefont {Miles~J}\
  \bibnamefont {Padgett}}, \ and\ \bibinfo {author} {\bibfnamefont {Erika}\
  \bibnamefont {Andersson}},\ }\bibfield  {title} {\enquote {\bibinfo {title}
  {Experimental high-dimensional two-photon entanglement and violations of
  generalized bell inequalities},}\ }\href@noop {} {\bibfield  {journal}
  {\bibinfo  {journal} {Nature Physics}\ }\textbf {\bibinfo {volume} {7}},\
  \bibinfo {pages} {677--680} (\bibinfo {year} {2011})}\BibitemShut {NoStop}%
\bibitem [{\citenamefont {Karimi}\ \emph {et~al.}(2014)\citenamefont {Karimi},
  \citenamefont {Boyd}, \citenamefont {de~la Hoz}, \citenamefont {de~Guise},
  \citenamefont {\ifmmode \check{R}\else \v{R}\fi{}eh\'a\ifmmode~\check{c}\else
  \v{c}\fi{}ek}, \citenamefont {Hradil}, \citenamefont {Aiello}, \citenamefont
  {Leuchs},\ and\ \citenamefont {S\'anchez-Soto}}]{PhysRevA.89.063813}%
  \BibitemOpen
  \bibfield  {author} {\bibinfo {author} {\bibfnamefont {E.}~\bibnamefont
  {Karimi}}, \bibinfo {author} {\bibfnamefont {R.~W.}\ \bibnamefont {Boyd}},
  \bibinfo {author} {\bibfnamefont {P.}~\bibnamefont {de~la Hoz}}, \bibinfo
  {author} {\bibfnamefont {H.}~\bibnamefont {de~Guise}}, \bibinfo {author}
  {\bibfnamefont {J.}~\bibnamefont {\ifmmode \check{R}\else
  \v{R}\fi{}eh\'a\ifmmode~\check{c}\else \v{c}\fi{}ek}}, \bibinfo {author}
  {\bibfnamefont {Z.}~\bibnamefont {Hradil}}, \bibinfo {author} {\bibfnamefont
  {A.}~\bibnamefont {Aiello}}, \bibinfo {author} {\bibfnamefont
  {G.}~\bibnamefont {Leuchs}}, \ and\ \bibinfo {author} {\bibfnamefont {L.~L.}\
  \bibnamefont {S\'anchez-Soto}},\ }\bibfield  {title} {\enquote {\bibinfo
  {title} {Radial quantum number of laguerre-gauss modes},}\ }\href {\doibase
  10.1103/PhysRevA.89.063813} {\bibfield  {journal} {\bibinfo  {journal} {Phys.
  Rev. A}\ }\textbf {\bibinfo {volume} {89}},\ \bibinfo {pages} {063813}
  (\bibinfo {year} {2014})}\BibitemShut {NoStop}%
\bibitem [{\citenamefont {Zhang}\ \emph {et~al.}(2018)\citenamefont {Zhang},
  \citenamefont {Qiu}, \citenamefont {Zhang},\ and\ \citenamefont
  {Chen}}]{zhang2018violation}%
  \BibitemOpen
  \bibfield  {author} {\bibinfo {author} {\bibfnamefont {Dongkai}\ \bibnamefont
  {Zhang}}, \bibinfo {author} {\bibfnamefont {Xiaodong}\ \bibnamefont {Qiu}},
  \bibinfo {author} {\bibfnamefont {Wuhong}\ \bibnamefont {Zhang}}, \ and\
  \bibinfo {author} {\bibfnamefont {Lixiang}\ \bibnamefont {Chen}},\ }\bibfield
   {title} {\enquote {\bibinfo {title} {Violation of a bell inequality in
  two-dimensional state spaces for radial quantum number},}\ }\href@noop {}
  {\bibfield  {journal} {\bibinfo  {journal} {Physical Review A}\ }\textbf
  {\bibinfo {volume} {98}},\ \bibinfo {pages} {042134} (\bibinfo {year}
  {2018})}\BibitemShut {NoStop}%
\bibitem [{\citenamefont {Walborn}\ \emph {et~al.}(2005)\citenamefont
  {Walborn}, \citenamefont {P{\'a}dua},\ and\ \citenamefont
  {Monken}}]{walborn2005conservation}%
  \BibitemOpen
  \bibfield  {author} {\bibinfo {author} {\bibfnamefont {SP}~\bibnamefont
  {Walborn}}, \bibinfo {author} {\bibfnamefont {S}~\bibnamefont {P{\'a}dua}}, \
  and\ \bibinfo {author} {\bibfnamefont {CH}~\bibnamefont {Monken}},\
  }\bibfield  {title} {\enquote {\bibinfo {title} {Conservation and
  entanglement of hermite-gaussian modes in parametric down-conversion},}\
  }\href@noop {} {\bibfield  {journal} {\bibinfo  {journal} {Physical Review
  A}\ }\textbf {\bibinfo {volume} {71}},\ \bibinfo {pages} {053812} (\bibinfo
  {year} {2005})}\BibitemShut {NoStop}%
\bibitem [{\citenamefont {Zhang}\ \emph {et~al.}(2016)\citenamefont {Zhang},
  \citenamefont {Prabhakar}, \citenamefont {Rosales-Guzm{\'a}n}, \citenamefont
  {Roux}, \citenamefont {Karimi},\ and\ \citenamefont
  {Forbes}}]{zhang2016hong}%
  \BibitemOpen
  \bibfield  {author} {\bibinfo {author} {\bibfnamefont {Yingwen}\ \bibnamefont
  {Zhang}}, \bibinfo {author} {\bibfnamefont {Shashi}\ \bibnamefont
  {Prabhakar}}, \bibinfo {author} {\bibfnamefont {Carmelo}\ \bibnamefont
  {Rosales-Guzm{\'a}n}}, \bibinfo {author} {\bibfnamefont {Filippus~S}\
  \bibnamefont {Roux}}, \bibinfo {author} {\bibfnamefont {Ebrahim}\
  \bibnamefont {Karimi}}, \ and\ \bibinfo {author} {\bibfnamefont {Andrew}\
  \bibnamefont {Forbes}},\ }\bibfield  {title} {\enquote {\bibinfo {title}
  {Hong-ou-mandel interference of entangled hermite-gauss modes},}\ }\href@noop
  {} {\bibfield  {journal} {\bibinfo  {journal} {Physical Review A}\ }\textbf
  {\bibinfo {volume} {94}},\ \bibinfo {pages} {033855} (\bibinfo {year}
  {2016})}\BibitemShut {NoStop}%
\bibitem [{\citenamefont {Miatto}\ \emph {et~al.}(2012)\citenamefont {Miatto},
  \citenamefont {di~Lorenzo~Pires}, \citenamefont {Barnett},\ and\
  \citenamefont {van Exter}}]{miatto2012spatial}%
  \BibitemOpen
  \bibfield  {author} {\bibinfo {author} {\bibfnamefont {Filippo~M}\
  \bibnamefont {Miatto}}, \bibinfo {author} {\bibfnamefont {H}~\bibnamefont
  {di~Lorenzo~Pires}}, \bibinfo {author} {\bibfnamefont {Stephen~M}\
  \bibnamefont {Barnett}}, \ and\ \bibinfo {author} {\bibfnamefont {Martin~P}\
  \bibnamefont {van Exter}},\ }\bibfield  {title} {\enquote {\bibinfo {title}
  {Spatial schmidt modes generated in parametric down-conversion},}\
  }\href@noop {} {\bibfield  {journal} {\bibinfo  {journal} {The European
  Physical Journal D}\ }\textbf {\bibinfo {volume} {66}},\ \bibinfo {pages}
  {1--11} (\bibinfo {year} {2012})}\BibitemShut {NoStop}%
\bibitem [{\citenamefont {Grenapin}\ \emph {et~al.}(2022)\citenamefont
  {Grenapin}, \citenamefont {Paneru}, \citenamefont {D'Errico}, \citenamefont
  {Grillo}, \citenamefont {Leuchs},\ and\ \citenamefont
  {Karimi}}]{grenapin2022super}%
  \BibitemOpen
  \bibfield  {author} {\bibinfo {author} {\bibfnamefont {Florence}\
  \bibnamefont {Grenapin}}, \bibinfo {author} {\bibfnamefont {Dilip}\
  \bibnamefont {Paneru}}, \bibinfo {author} {\bibfnamefont {Alessio}\
  \bibnamefont {D'Errico}}, \bibinfo {author} {\bibfnamefont {Vincenzo}\
  \bibnamefont {Grillo}}, \bibinfo {author} {\bibfnamefont {Gerd}\ \bibnamefont
  {Leuchs}}, \ and\ \bibinfo {author} {\bibfnamefont {Ebrahim}\ \bibnamefont
  {Karimi}},\ }\bibfield  {title} {\enquote {\bibinfo {title} {Super-resolution
  enhancement in bi-photon spatial mode demultiplexin},}\ }\href@noop {}
  {\bibfield  {journal} {\bibinfo  {journal} {arXiv preprint arXiv:2212.10468}\
  } (\bibinfo {year} {2022})}\BibitemShut {NoStop}%
\bibitem [{\citenamefont {Valencia}\ \emph {et~al.}(2021)\citenamefont
  {Valencia}, \citenamefont {Srivastav}, \citenamefont {Leedumrongwatthanakun},
  \citenamefont {McCutcheon},\ and\ \citenamefont
  {Malik}}]{valencia2021entangled}%
  \BibitemOpen
  \bibfield  {author} {\bibinfo {author} {\bibfnamefont {Natalia~Herrera}\
  \bibnamefont {Valencia}}, \bibinfo {author} {\bibfnamefont {Vatshal}\
  \bibnamefont {Srivastav}}, \bibinfo {author} {\bibfnamefont {Saroch}\
  \bibnamefont {Leedumrongwatthanakun}}, \bibinfo {author} {\bibfnamefont
  {Will}\ \bibnamefont {McCutcheon}}, \ and\ \bibinfo {author} {\bibfnamefont
  {Mehul}\ \bibnamefont {Malik}},\ }\bibfield  {title} {\enquote {\bibinfo
  {title} {Entangled ripples and twists of light: radial and azimuthal
  laguerre--gaussian mode entanglement},}\ }\href@noop {} {\bibfield  {journal}
  {\bibinfo  {journal} {Journal of optics}\ }\textbf {\bibinfo {volume} {23}},\
  \bibinfo {pages} {104001} (\bibinfo {year} {2021})}\BibitemShut {NoStop}%
\bibitem [{\citenamefont {Bolduc}\ \emph {et~al.}(2013)\citenamefont {Bolduc},
  \citenamefont {Bent}, \citenamefont {Santamato}, \citenamefont {Karimi},\
  and\ \citenamefont {Boyd}}]{bolduc2013holo}%
  \BibitemOpen
  \bibfield  {author} {\bibinfo {author} {\bibfnamefont {E.}~\bibnamefont
  {Bolduc}}, \bibinfo {author} {\bibfnamefont {N.}~\bibnamefont {Bent}},
  \bibinfo {author} {\bibfnamefont {E.}~\bibnamefont {Santamato}}, \bibinfo
  {author} {\bibfnamefont {E.}~\bibnamefont {Karimi}}, \ and\ \bibinfo {author}
  {\bibfnamefont {R.~W.}\ \bibnamefont {Boyd}},\ }\bibfield  {title} {\enquote
  {\bibinfo {title} {Exact solution to simultaneous intensity and phase
  encryption with a single phase-only hologram},}\ }\href {\doibase
  10.1364/OL.38.003546} {\bibfield  {journal} {\bibinfo  {journal} {Optics
  Letters}\ }\textbf {\bibinfo {volume} {38}},\ \bibinfo {pages} {3546--3549}
  (\bibinfo {year} {2013})}\BibitemShut {NoStop}%
\end{thebibliography}%

\vspace{0.5cm}
\vspace{1 EM}

\noindent\textbf{Acknowledgments}
\noindent The authors would like to acknowledge Dr. Yingwen Zhang for providing the LabView code to analyse the data extracted from the time-stamping camera. This work was supported by Canada Research Chairs (CRC), Canada First Research Excellence Fund (CFREF) Program, NRC-uOttawa Joint Centre for Extreme Quantum Photonics (JCEP) via High Throughput and Secure Networks Challenge Program at the National Research Council of Canada. DZ and FS acknowledge support by the ERC Advanced Grant QU-BOSS (grant agreement no. 884676) and "bando per il finanziamento di progetti di ricerca congiunti ed individuali per la mobilità all'estero di studenti di dottorato" issued by Sapienza.
\vspace{1 EM}

\noindent\textbf{Author Contributions}
A.D conceived the idea. D.Z., N.D., and A.D. devised the experimental setup. D.Z., with the help of N.D. and A.D. performed the experiment. D.Z. analyzed the data. N.D. analyzed the spatial mode correlations. E.K. and F.S. supervised the project. D.Z., N.D., and A.D. prepared the first version of the manuscript. All authors contributed to the writing of the manuscript.
\vspace{1 EM}

\noindent\textbf{Author Information}
\noindent The authors declare no competing financial interests. Correspondence and requests for materials should be addressed to aderrico@uottawa.ca.
\clearpage
%
\renewcommand{\figurename}{\textbf{Figure}}
\setcounter{figure}{0} \renewcommand{\thefigure}{\textbf{S{\arabic{figure}}}}
\setcounter{table}{0} \renewcommand{\thetable}{S\arabic{table}}
\setcounter{section}{0} \renewcommand{\thesection}{Section~\arabic{section}}

\section*{Methods}

\subsection{Detailed experimental setup.}

A Gaussian beam with a wavelength of 405 nm is produced through the second harmonic generation of a 810 nm pulsed Ti:Sa laser (Chameleon Vision II), the latter has a pulse duration of 150 fs and a repetition rate of 80 MHz. The beam is magnified to $\approx$ 1 cm beam waist and sent in the input of a Michelson interferometer. A reflective liquid crystal spatial light modulator is placed in one arm of the interferometer. At the interferometer output, all the diffraction orders of the SLM, except for the first, are filtered out by a slit placed in the SLM's Fourier plane. The slit allows keeping the reference beam (corresponding to the beam going through the interferometer arm without the SLM) with a different transverse wavevector, thus allowing to perform off-axis digital holography and, at the same time, keeping good interference stability. The phase masks applied on the SLM allows the generation of arbitrary optical fields by means of the technique introduced in Ref. \cite{bolduc2013holo}.   After filtering, both beams are collimated and sent on the 0.5 mm thick type-I BBO crystal for SPDC generation. The vertically polarized down-converted light is collimated by a lens of focal length $f=25$ cm, split in two copied by a sequence consisting of a half-wave plate rotated by 22.5° and a polarizing beam splitter (effectively working as an ordinary beam splitter). The two copies are sent on parallel paths (by sending them on another  polarizing beam splitter and changing the polarizations in such a way as to maximize the intensity on one output port). The two beams have a lateral shift, that avoids them from being overlapped, and are focused, by means of a  $f=50$ cm lens, on the TPX3CAM sensor. In front of the sensor, a 3 nm bandpass filter is applied to ensure the frequency degeneracy of the analyzed photons. In the far field configuration, used to reconstruct the phase-matching function, an additional $f = 20$ cm lens is placed in front of the camera in a confocal configuration.

\subsection{Data acquisition and analysis.}
Data were acquired by collecting SPDC light on the TPX3CAM for 1 minute for each data set. We collected data for cases where both the reference and pump beam were sent on the BBO crystal and cases where only the pump beam was sent on the crystal, after blocking the reference arm. The acquired data files report the timestamp at which counts were detected (for more information see Refs. \cite{fisher2016timepixcam,zhang2021high,gao2022high,zhang2022ray}). In our case, since we shine two copies of the SPDC light on different regions on the camera, we can have counts from these regions detected in the same time window. We considered as coincidences the counts from the two regions with a timestamp difference of 5 ns. From this set of counts, we analyzed the spatial correlations, confirming the  validity of the thin crystal approximation. A weak constant background in the correlation plot is always present (due to dark counts and background light) and can be reduced by removing the counts outside the correlation region. Coincidence images were obtained by plotting the positions of the counts selected as coincidences.

The resulting coincidence images were analyzed using standard off axis digital holography analysis \cite{yamaguchi2006phase}. In off-axis digital holography, with $\mathcal{E}_\text{ref}(x,y)=A \exp(-(x^2+y^2)/w_r^2)\exp(i 2\pi (x+y)/\Lambda)$, one has 
\begin{align}
    \abs{\mathcal{E}_{ref}+\mathcal{E}_p}^2=&\abs{\mathcal{E}_\text{ref}}^2+\abs{\mathcal{E}_p}^2\nonumber \\
&+2Ae^{-\frac{r^2}{w_r^2}}(\mathcal{E}_pe^{-i2\pi \frac{(x+y)}{\Lambda}}+c.c.).\nonumber
\end{align}
From a spatial Fourier transform one can hence isolate the term proportional to $\mathcal{E}_p$ and reconstruct the amplitude and phase of the unknown field.

The analysis of the reconstructed states in terms of OAM, HG, and LG modes has been conducted by direct calculations of the expansion coefficients in the respective bases. The errors on the Fidelities have been obtained by repeating the analysis for different state reconstructions, where the original coincidence images were modified pixel by pixel by random amounts within the uncertainty, given by the square root of the coincidences assuming Poissonian statistics. In the main text, we report the average Fidelity and the standard deviation over 20 different realizations.

\subsection{Parity conservation.}

The coefficients $C_{m_i,m_i}^{m_s,n_s}$ in the HG expansion of the SPDC state are given by the integral:
\begin{align}
    C_{m_i,m_i}^{m_s,n_s}:=(\bra{m_i,n_i}&\otimes\bra{m_s,n_s})\ket{\Psi}=\nonumber\\ \mathcal{N}\iint&\mathcal{E}(x,y)e^{-2(\frac{x}{w_p})^2}h_{m_i}\left(\frac{x}{w_p}\right)\,h_{m_s}\left(\frac{x}{w_p}\right)\nonumber\\
\times &e^{-2(\frac{y}{w_p})^2}h_{n_i}\left(\frac{y}{w_p}\right)h_{n_s}\left(\frac{y}{w_p}\right) dx\,dy.\nonumber
\end{align}
We consider the case in which $\mathcal{E}(x,y)=F_x(x)\,F_y(y)$ where $F_x$ and $F_y$ are even or odd functions of $x$ and $y$, respectively (which is the case if the pump is in a HG mode). We have
$C_{m_i,m_i}^{m_s,n_s}=I_{F_x}\times I_{F_y}$, where $$I_{F_{\xi}}:=\sqrt{\mathcal{N}}\int_{-\infty}^{\infty}e^{-2(\frac{\xi}{w_p})^2}F_{\xi}(\xi)h_{l_i}\left(\frac{\xi}{w_p}\right)h_{l_s}\left(\frac{\xi}{w_p}\right)d\xi$$
with $l=m,n$ for $\xi=x,y$, respectively. The product $h_{l_i}\left(\frac{\xi}{w_p}\right)h_{l_s}\left(\frac{\xi}{w_p}\right)$ is even/odd if $l_i+l_s$ is even/odd. Thus, the integral is zero if the parity of $F_{\xi}$ is different than the parity of $l_i+l_s$, hence the conservation law mentioned in the main text. 
\end{document}